\documentclass[fleqn,usenatbib]{mnras}

\usepackage{newtxtext,newtxmath}
\usepackage{bm}
\usepackage{physics}
\usepackage{microtype}

\usepackage[T1]{fontenc}

\DeclareRobustCommand{\VAN}[3]{#2}
\let\VANthebibliography\thebibliography
\def\thebibliography{\DeclareRobustCommand{\VAN}[3]{##3}\VANthebibliography}

\usepackage{graphicx}	
\usepackage{amsmath}	

\title[Effects of Viscosity on Sloshing Cold Fronts in Galaxy Clusters]{Effects of Viscosity on Sloshing Cold Fronts in Galaxy Clusters}

\author[Hsieh, Ming-Hsueh et al.]{
Ming-Hsueh Hsieh,$^{1}$\thanks{E-mail: arthur20010514@gmail.com}
H.-Y. Karen Yang,$^{1,2,3}$\thanks{E-mail: hyang@phys.nthu.edu.tw}
and John ZuHone$^{4}$
\\
$^{1}$Institute of Astronomy, National Tsing Hua University, Hsinchu 30013, Taiwan (R.O.C.)\\
$^{2}$Center for Informatics and Computation in Astronomy, National Tsing Hua University, Hsinchu 30013, Taiwan (R.O.C.)\\
$^{3}$Physics Division, National Center for Theoretical Sciences, Taipei 106017, Taiwan (R.O.C.)\\
$^{4}$Center for Astrophysics | Harvard \& Smithsonian, 60 Garden Street, Cambridge, MA 02138, USA
}

\date{Accepted XXX. Received YYY; in original form ZZZ}

\pubyear{2024}

\begin{document}
\label{firstpage}
\pagerange{\pageref{firstpage}--\pageref{lastpage}}
\maketitle

\begin{abstract}

The viscous properties of the intracluster medium (ICM) remain poorly constrained. Cold fronts—sharp discontinuities formed during cluster mergers—offer a potential avenue to probe the effective viscosity of the ICM. Velocity shear across these fronts should generate Kelvin–Helmholtz instabilities (KHI), unless viscosity or magnetic tension suppresses them. We perform cluster merger simulations incorporating four ICM viscosity models: (A) inviscid, (B) isotropic Spitzer viscosity, (C) anisotropic Braginskii viscosity, and (D) Braginskii viscosity limited by microinstabilities. The isotropic Spitzer viscosity (case B) strongly suppresses KHI, producing smooth cold front surfaces, while the inviscid (A) and microinstability-limited (D) cases show prominent ripples. The Braginskii case (C) yields intermediate suppression. We also vary the plasma $\beta$ parameter ($\beta \approx$ 100 and 1600) to examine how a changing magnetic field strength affects the results. Stronger magnetic fields further suppress KHI, leading to smoother fronts and reduced differences between different viscosity models, while also widening the range of permitted pressure anisotropies when microinstability-based limiters are present. These results indicate that both viscosity and magnetic fields play crucial roles in stabilising sloshing cold fronts in galaxy clusters.

\end{abstract}




\section{Introduction}

X-ray observations have revealed valuable information about the intracluster medium (ICM), the hot plasma filling the dark matter halos of galaxy clusters. 
Although it is known that the ICM is a weakly collisional, magnetized plasma, key microphysical properties such as the strength and configuration of the magnetic field, as well as the effective viscosity, remain poorly constrained. 
In particular, it is unclear whether the viscosity in the ICM is isotropic or anisotropic, and how it combines with magnetic fields to suppress hydrodynamic instabilities such as the Kelvin–Helmholtz instability (KHI). Numerical simulations provide a powerful tool for probing these unknowns and testing theoretical models of ICM microphysics \citep{Berlok2019,Marin-Gilabert2025}.

X-ray observations of galaxy clusters often show sharp surface brightness discontinuities. Spectral analysis of the emission on either side of these images has revealed that in most cases the brighter side of the edge is the colder side \citep{Markevitch_2000,Vikhlinin_2001}. Brighter regions correspond to denser areas since the X-ray surface brightness depends on the square of the density. Hence, these jumps in gas density have been called “cold fronts” \citep[for a detailed review, see][]{Markevitch(2007)}. Observations indicate that cold fronts can form during major cluster mergers; a prominent example is the "Bullet Cluster" \citep{Markevitch(2002)}.
However, cold fronts are also observed in relaxed clusters that do not exhibit evidence of major mergers. In these cases, the cold fronts are thought to arise from sloshing motions from encounters with small infalling groups or subclusters that gravitationally perturb the cluster core \citep{Ascasibar_2006, Tittley_2005, ZuHone_2010}. During a minor merger or gravitational perturbation, a subcluster passes near the core of a larger galaxy cluster. This encounter displaces the dense, low-entropy gas at the center of the cluster from the minimum of the gravitational potential. Unlike the collisionless dark matter, the intracluster gas is subject to ram pressure from the surrounding medium and is temporarily pushed outward. Once the external perturbation subsides, the gravitational potential of the cluster pulls the displaced gas back toward the center. Due to the conservation of angular momentum and the asymmetry of the initial displacement, the gas does not return in a straight line. Instead, it begins to oscillate—or "slosh"—in the gravitational potential well. These motions generate spiral-like patterns and create sharp contact discontinuities in density and temperature, observed as cold fronts in X-ray surface brightness maps.

The KHI is a type of fluid instability in a shear layer, which causes the interface of the layer to become unstable to small perturbations and form structures such as ripples. At the interfaces of cold fronts in clusters, the velocity difference between the inner part and the outer part of the interface produces a shear which is caused by the sloshing motions. This velocity shear causes the interface to become susceptible to the KHI. Early X-ray observations showed that the cold front interfaces typically appeared smooth \citep{Markevitch_2000, Vikhlinin_2001} and free from the ripples which would serve as the evidence of KHI. However, with the benefit of deeper exposures providing better statistics, several cold fronts have been found to exhibit evidence of the KHI, such as ripple-like or eddy structures \citep{Werner_2016,Su_2017,Wang_2018,Walker2017}. More observational data are still required to fully determine the morphology of cold fronts. However, 
even in cases where eddy-like structures are observed, the growth of the KHI appears to be partially suppressed to some extent (with an estimated effective viscosity of $\sim 10\%$ of the Spitzer value in \citealt{Wang_2018} and $\sim 5\%$ in \citealt{Werner_2016}). Previous simulations \citep{ZuHone(2011), Roediger_2013, ZuHone_2015, Zuhone2016} suggest that both viscosity and magnetic fields can play a significant role in suppressing the KHI in sloshing cold fronts. These works explore different physical scenarios: \citet{ZuHone(2011)} focuses on the influence of varying initial plasma $\beta$ values, while \citet{ZuHone_2015} examines the effects of different viscosity models on the development of cold front structures. However, it is not clear which mechanisms are more dominant in suppressing the KHI.\footnote{{Simulations including anisotropic thermal conduction \citep{ZuHone_2012} also found that conduction could indirectly aid in suppressing the KHI by modifying the gas density distribution below the cold front. However, in this work we focus on the effect of viscosity.}}

In the ICM, Coulomb collisions between ions and electrons are relatively infrequent due to the low density and high temperature, resulting in a large mean free path. Under such weakly collisional conditions, the classical Spitzer formulation \citep{Spitzer(1962)} becomes invalid. In addition, the presence of magnetic fields introduces anisotropy in transport processes, since the Larmor radii of the electrons and ions are many orders of magnitude smaller than the Coulomb mean free paths. In such a magnetized plasma, momentum and heat are transported more efficiently along magnetic field lines than across them, leading to Braginskii viscosity, an anisotropic form of viscous stress originating from pressure anisotropies in the weakly collisional, magnetized medium \citep{Braginskii(1965), Schekochihin(2005),Kunz2012,Squire2023}. Furthermore, recent plasma physics simulations have shown that Larmor-scale kinetic instabilities, such as the firehose and mirror instabilities, would be triggered and act to regulate pressure anisotropy and suppress viscosity in the ICM \citep{Kunz(2014)}. How these microphysics instabilities impact the cold front structures still remain unsolved. Inspired by them, we attempt to understand the roles of microphysics in cluster mergers through numerical simulations. 

Our simulations aim to explore the nature of the ICM viscosity in the context of the cores of relaxed galaxy clusters, where sloshing cold fronts are found. By incorporating different viscosity models and magnetic field strengths, we compare the simulated cold fronts with observations, seeking to provide insights and predictions for future research.

This paper is organized as follows.
\begin{itemize}
    \item Section \ref{sec:Methods} describes the numerical setup and simulation methods, including the initial conditions, viscosity models, and the governing equations.
    \item Section \ref{sec:Results} presents the simulation results, focusing on how different viscosity models and magnetic field strengths affect the stability and morphology of sloshing cold fronts.
    \item Section \ref{sec:Discussion} discusses the results and compares them with previous studies.
    \item Section \ref{sec:Conclusions} presents a summary of this work.
\end{itemize}
\section{Methods}
\label{sec:Methods}

We conduct 3D Braginskii-magnetohydrodynamic (MHD) simulations of binary cluster mergers using FLASH, an adaptive-mesh-refinement (AMR) MHD code. The algorithms employed in FLASH are detailed in \citet{Fryxell(2000)} and \citet{Dubey2009}. Our cluster merger simulations incorporated four different viscosity models: (A) no viscosity, (B) isotropic Spitzer viscosity, (C) Braginskii viscosity, and (D) Braginskii viscosity limited by microscopic plasma instabilities. These assumptions follow the approaches outlined in \cite{Kingsland(2019)} and the equations are provided in Section~\ref{subsec:2.1}. The cluster merger scenario follows that of \cite{ZuHone(2011)} and will be described in detail in Section~\ref{subsec:2.3}. For each of the four assumptions, we also introduce varying magnetic energy to investigate how magnetic field strength influences the impact of viscosity on the cold fronts and sloshing motions.

\subsection{Equations}
\label{subsec:2.1} 
In our cluster merger simulations, we solve the Braginskii-MHD equations as follows:
\begin{equation}
    \frac{\partial\rho}{\partial t} + \div{(\rho \bm{v})} = 0
    \label{f1}
\end{equation}
\begin{equation}
    \frac{\partial(\rho \bm{v})}{\partial t} +  \div{(\rho \bm{v} \bm{v} - \frac{\vb{B}\vb{B}}{4\pi} +  \ p_{\rm tot}\vb{I})} = \rho \vb{g} - \div{\Pi} 
    \label{f2}
\end{equation}
\begin{equation}
    \frac{\partial E}{\partial t}+
    \div{[
    \bm{v}(E+p_{\rm tot})-
    \frac{\vb{B}(\bm{v} \cdot\vb{B})}{4\pi}
    ]}=
    \rho\vb{g}\cdot\bm{v}-
    \div{(\Pi\cdot\bm{v})}
    \label{f3}
\end{equation}
\begin{equation}
    \frac{\partial\vb{B}}{\partial t} + \div{(\bm{v}\vb{B}-\vb{B}\bm{v})} = 0,
    \label{f4}
\end{equation}
where $\rho$ is gas density, $\bm{v}$ is gas velocity, $E$ is the total energy density, $\vb{B}$ is the magnetic field, $\vb{I}$ is the identity tensor, and $p_{\rm tot}=p+p_{\rm B}=p+B^2/(8\pi)$ is the total (thermal plus magnetic) pressure.  
In this work, we consider several models for the viscosity tensor $\Pi$. For the case of isotropic viscosity, we adopt the Spitzer viscosity \citep{Spitzer(1962)}, appropriate for a collisional, non-magnetized, fully ionized plasma. The isotropic viscosity tensor is defined as
\begin{equation}
    \Pi_{\rm iso} = -\mu\left[\nabla{\bf v}+(\nabla{\bf v})^\top-\frac{2}{3}\nabla\cdot{\bf v}\right],
    \label{f5}
\end{equation}
where $\mu$ is the dynamic viscosity coefficient
\begin{equation}
    \mu=2.2\times10^{-15} \frac{T^{5/2}}{\ln{\Lambda}}\, {\rm g}\, {\rm cm}^{-1}\, {\rm s}^{-1},
    \label{f7}
\end{equation}
where $\ln{\Lambda=40}$ is the Coulomb logarithm following the approximation of ICM conditions \citep[e.g.,][]{Roediger_2013}.

For the anisotropic Braginskii viscosity, it can be expressed as \citep{Braginskii(1965)}:
\begin{equation}
    \Pi_{\rm aniso}=-3\mu(\vb{b}\vb{b}-\frac{1}{3}\vb{I})(\vb{b}\vb{b}-\frac{1}{3}\vb{I}):\nabla\bm{v}, 
    \label{f6}
\end{equation}
where $\vb{b}$ is the unit vector of the magnetic field.  
For non-magnetized, collisional fluids, the isotropic Spitzer viscosity is a good approximation. However, in the magnetized, weakly collisional ICM, the gyro-radii of charged particles are very small compared to the collisional mean free path of the particles, which leads to anisotropic or Braginskii viscosity. In essence, the Braginskii viscosity arises from pressure anisotropy due to the conservation of the adiabatic invariants of charged particles within the magnetized plasma, which is balanced by its relaxation via ion-ion collisions \citep{Schekochihin(2005), Schekochihin(2006),Arzamasskiy2023}. This relationship is described by
\begin{equation}
    \delta p = p_{\perp}-p_{\parallel} = 0.96\frac{p_{\rm i}}{\bm{\nu}_{\rm ii}}\dv{}{t}\ln{\frac{B^3}{\rho^2}}=3\mu(\vb{b}\vb{b}-\frac{1}{3}\vb{I}):\nabla\bm{v},
    \label{f8}
\end{equation}
where $\nu_{\rm ii}$ is the ion-ion collision frequency and $p_{\rm i}$ is thermal pressure for ions. We apply both isotropic and anisotropic models of viscosity in our simulations to determine which fares better in comparison to the observations.

Plasma physics simulations have found that Larmor-scale instabilities (firehose instability and mirror instability) would be triggered and act to regulate the the pressure anisotropy \citep{Sharma2006,Kunz(2014),St-Onge2020}. 
They found that the pressure anisotropy should be limited within the microinstability thresholds:
\begin{equation}
    -\frac{2}{\beta}\leq\Delta_{\rm p}\equiv\frac{\delta p}{p}\leq\frac{1}{\beta},
    \label{f9}
\end{equation}
where $\beta = p/p_{\rm B}$ is the plasma $\beta$ parameter, indicating the ratio of thermal to magnetic pressures. The left-hand side of equation \ref{f9} corresponds to the firehose instability \citep[$\Delta_{\rm p}<-\frac{2}{\beta}$;][]{Rosenbluth1956,Chandrasekhar1958,Parker1958}, and the right-hand side corresponds to the mirror instability \citep[$\Delta_{\rm p}>\frac{1}{\beta}$]{Barnes1966,Hasegawa1969,Southwood1993,Hellinger2007}. The firehose instability occurs when the pressure along the magnetic field lines is greater than the pressure perpendicular to the field to such an extent that it overcomes the restoring force of the magnetic tension, leading to the unstable growth of Alfv\'enic perturbations. In contrast, the mirror instability occurs when the pressure perpendicular to the magnetic field lines is greater than the pressure parallel to the field by an extent that exceeds the magnetic pressure. In this case, the plasma will expand sideways, distorting the magnetic field into local magnetic ``bottles''. These bottle structures trap particles which bounce between mirror structures on either side. At the ends of the bottles, particles can escape due to adiabatic invariance, lowering the thermal pressure and causing the magnetic field lines to further contract, reinforcing the instability. In either case, the fastest-growing modes of the instability are near the Larmor scale \citep{Kunz(2014),Melville2016}. These modes efficiently pitch-angle scatter particles, reducing the anisotropy and restoring marginal stability. This effect acts as a feedback mechanism to regulate the anisotropy. In the weakly collisional, magnetized ICM, the high plasma-$\beta$ conditions are likely to trigger both firehose and mirror instabilities. Therefore, in one of our models, we impose constraints on pressure anisotropy to account for the regulation imposed by these microphysical instabilities. For clarity, Equations \ref{f10} and \ref{f11} express the anisotropic pressure with limiters and the anisotropic viscosity with pressure limiters in terms of $\Delta_{\rm p}$. \footnote{We note that there are other relevant bounds for the pressure anisotropy in plasmas such as the ICM due to the oblique firehose instability \citep{Hellinger2000,Hellinger2008,Bott2021} or ion-cyclotron instability \citep{Gary1997,Ley_2024}. For simplicity, we do not include these bounds in the current study, but will include them in future work.}

\begin{equation}
    \Delta_{\rm p, lim}=
    \begin{cases}
        \frac{-2}{\beta}, ~\rm  ~\Delta_{\rm p} \leq\frac{-2}{\beta}\\
        \Delta_{\rm p}, ~\rm \frac{-2}{\beta}<\Delta_{\rm p} <\frac{1}{\beta}\\
        \frac{1}{\beta}, ~\rm ~\Delta_{\rm p} \geq\frac{1}{\beta}\\
    \end{cases}
    \label{f10}
\end{equation}
\begin{equation}
    \Pi_{\rm aniso, ~limiter}=-(\vb{b}\vb{b}-\frac{1}{3}\vb{I})p\Delta_{\rm p, lim}
    \label{f11}
    \end{equation}

\subsection{Initial conditions}
\label{subsec:2.3}
We simulate the merger of two galaxy clusters, following the setup by \cite{ZuHone(2011)}. The primary cluster is modeled as a cool-core cluster with $ M \sim 1.25\times 10^{15} M_\odot$, while the mass of the sub-cluster is $2.5\times 10^{14} M_\odot$. For the dark matter profile of the cluster, the \cite{Hernquist(1990)} profile is utilized:
\begin{equation}
    \rho_{\rm DM}(r)=\frac{M_{\rm tot}}{2\pi a^3}\frac{1}{(r/a)(1+r/a)^3}.
\end{equation}
Here, $M_{\rm tot}$ is the total mass, and $a=600$ kpc is the scale length of the dark matter halo.

The sub-cluster is initially positioned at a distance $d$ = 3 Mpc from the main cluster at the beginning of the simulations. The initial impact parameter is $b$ = 0.5 Mpc, and the relative velocity of sub-cluster is 1466 km s$^{-1}$ at the beginning of the simulation, resulting in the total kinetic energy being half of its potential energy. The two clusters come closest around $t$ = 1.5 Gyr, and after this epoch, the sloshing motions begin. In Section \ref{sec:Results}, we present data at 2.2 Gyr and 4 Gyr to compare the physical parameters after the core passage of the merger.

The simulation box is a cube with a side length of 8 Mpc, refined using AMR in order to reduce computational expense. Our primary focus lies in the cluster center, where sloshing motions predominantly occur. The refinement criteria in our simulations are based on gradients of gas density and temperature, and refinement within a fixed radius of 150 kpc from the main cluster center. 
The coarsest cell size on our AMR grid is $\Delta x = 62.5$ kpc and the finest cell size is $\Delta x = 3.9$ kpc. The scales of the sloshing motions are approximately 50-200~kpc. Thus, this resolution is sufficient for analyzing the sloshing motions in different viscosity environments.

In our simulations, we utilize the kinematic viscosity $(\nu=\mu/\rho)$ to compute the simulation time step. 
With an explicit solver for viscosity, the simulation time step must obey the inequality $\Delta t\leq \frac{1}{2} \frac{ \Delta x^2}{\nu}$, where $\Delta x$ is the grid size. Higher kinematic viscosity leads to smaller time steps and significantly increases the computational time. To address this issue, we impose an upper limit on the kinematic viscosity, such that $\nu \leq 10^{31}\,\mathrm{cm}^2\,\mathrm{s}^{-1}$. This ceiling is somewhat arbitrary but necessary for maintaining computational efficiency. In our simulations, $\nu$ typically ranges from $\sim 10^{28}$ to $10^{31}\,\mathrm{cm}^2\,\mathrm{s}^{-1}$ in the central region where our analyses are conducted. Given this range, we expect that the imposed ceiling does not significantly affect the overall results.

The simulation suite comprises four models of ICM viscosity. The first model does not involve any viscosity. The second model includes isotropic Spitzer viscosity, known for effectively suppressing KHI \citep{Roediger2013a,Roediger_2013}, though it might considered too extreme for some of the observed cold fronts \citep{Werner_2016}. The third model incorporates Braginskii viscosity, expected to operate in magnetized plasmas, in which momentum diffusion occurs predominantly along the magnetic field lines. The fourth model combines Braginskii viscosity with microinstabilities, introducing effects that limit pressure anisotropies. In Section \ref{sec:Results}, we present and compare the results of the four models: (A) no viscosity, (B) isotropic viscosity, (C) anisotropic viscosity, and (D) anisotropic viscosity limited by the microinstabilities.

For the initial magnetic field, we follow \cite{ZuHone(2011)} and \cite{ZuHone_2013} to set up a tangled magnetic field with a power-law spectrum ranging from 500 kpc to 43 kpc. In each of the four viscosity models, we introduce two levels of magnetic energy in the ICM to investigate the influence of magnetic field strength on cluster mergers. To facilitate interpretation, we employ the plasma $\beta$ parameter to quantify the amount of magnetic energy in the simulations. In the ICM, the plasma $\beta$ typically has a value between $\sim$100-1000 due to the weak magnetic field. 
In this paper, we use two values of plasma $\beta$ ($\beta = 100$ and $\beta = 1600$) to examine the potential impact of the initial magnetic field strength on the sloshing cold fronts. The magnetic field plays an important role, because magnetic tension can suppress the KHI on its own \citep{ZuHone(2011)}. As already noted above, the magnetic field makes the viscosity anisotropic, and a stronger field strength widens the microinstability bounds. The setup of our simulations is summarized in Table \ref{table1}.
\begin{table}
	\centering
	\caption{Eight simulations for different initial plasma $\beta$ ($\beta$) and different assumptions for ICM viscosity.}
	\label{table1}
	\begin{tabular}{lccr} 
		\hline
		Simulation & Initial $\beta$ & Assumption for ICM viscosity\\
		\hline
		A100&100&no viscosity\\
		A1600&1600&no viscosity\\
		B100&100&isotropic viscosity\\
            B1600&1600&isotropic viscosity\\
            C100&100&anisotropic viscosity\\
		C1600&1600&anisotropic viscosity\\
		D100&100&anisotropic viscosity limited by microinstabilities\\
            D1600&1600&anisotropic viscosity limited by microinstabilities\\
		\hline
	\end{tabular}
\end{table}

\section{Results}
\label{sec:Results}
In this section, we present our simulation results. All simulations discussed in Section \ref{subsec:3.1} and \ref{subsec:3.2} are initialized with $\beta \sim 1600$. Section \ref{subsec:3.1} illustrates the evolution of the main cluster during the cluster merger. We delve into the effects of viscosity in detail in Section \ref{subsec:3.2}. Finally, Section \ref{subsec:3.3} presents simulations with $\beta \sim 100$. 
Our primary focus in this work is to investigate how the sloshing cold front influences the central part of the main cluster, and we present zoom-in figures to illustrate this effect. 

\subsection{Overall evolution}
\label{subsec:3.1}

The entire simulation duration for the cluster merger spans $\sim$ 5 Gyr. Although the two clusters come closest around $t=1.5$ Gyr, it takes time for the cold front structures to form in the central part of the main cluster. We have chosen the epochs of 2.2 Gyr and 4.0 Gyr to represent the early and later phases of cold front formation, respectively. 

\begin{figure}
 \includegraphics[width=\columnwidth]{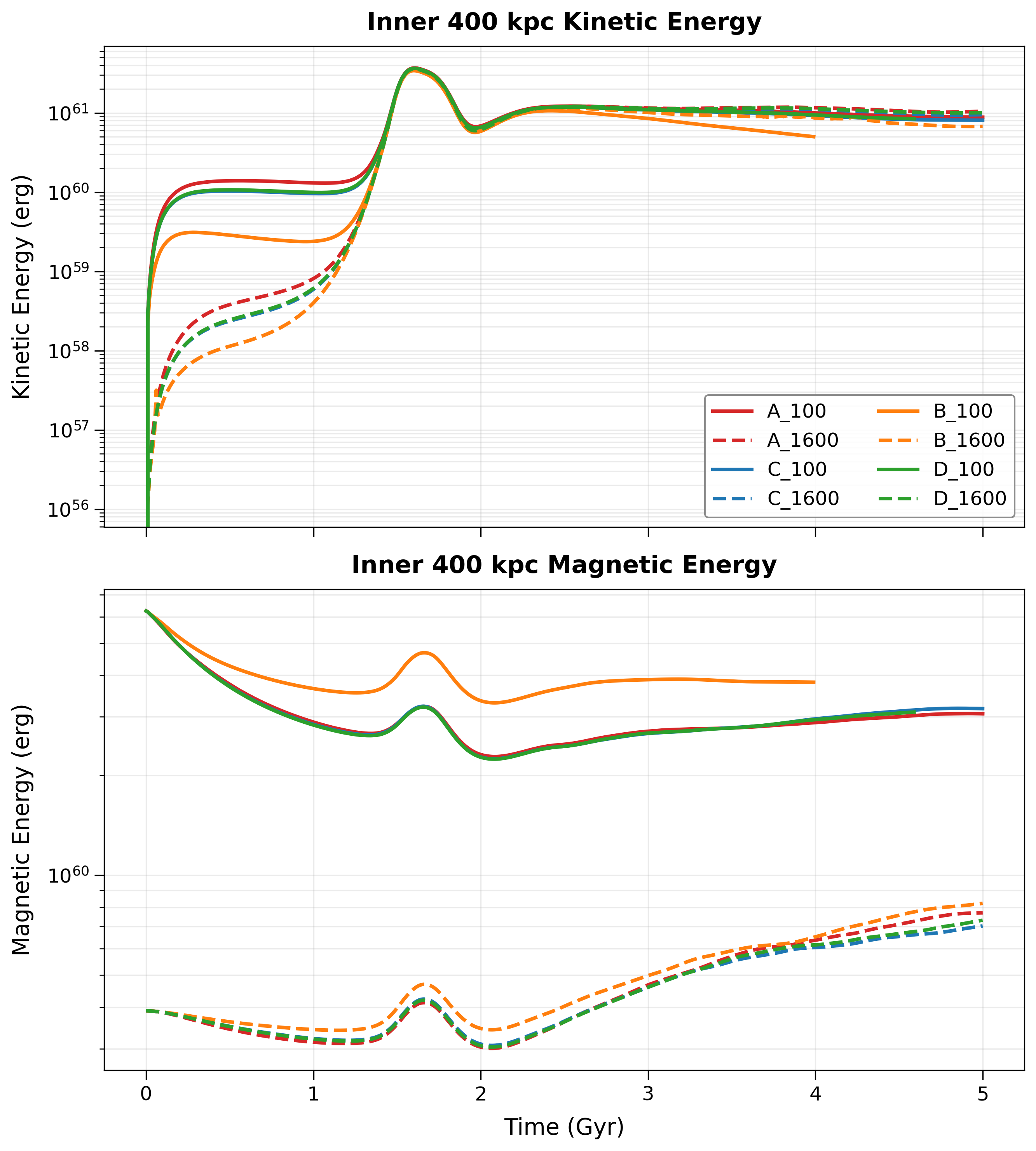}
 \caption{Time evolution of the kinematic energy (top panel) and the magnetic energy (bottom panel) within the inner region ($r < 400$ kpc). In the early stages, the magnetic energy decreases as it is converted into kinetic energy.}
 \label{evolution_E}
\end{figure}

Figure \ref{evolution_E} presents the time evolution of the kinetic (top panel) and magnetic (bottom panel) energies within the inner region ($r < 400$ kpc). 
During the initial stage, the initially tangled magnetic fields unwind and do work on the surrounding gas (since they are not in general force-free), thereby converting a fraction of the magnetic energy into kinetic energy. However, viscosity acts to damp these magnetically induced gas motions. Consequently, the various simulated cases exhibit slightly different time dependencies in energy depending on their specific viscous treatments. Notably, Case B (isotropic viscosity) exhibits distinct evolution because its strong viscous forces severely restrict the gas motions required for the magnetic fields to unwind. As a result, across both $\beta$ regimes, Case B retains higher magnetic energy and produces lower kinetic energy during the early stages. This effect is particularly pronounced in the low-$\beta$ cases.

\begin{figure}
 \includegraphics[width=\columnwidth]{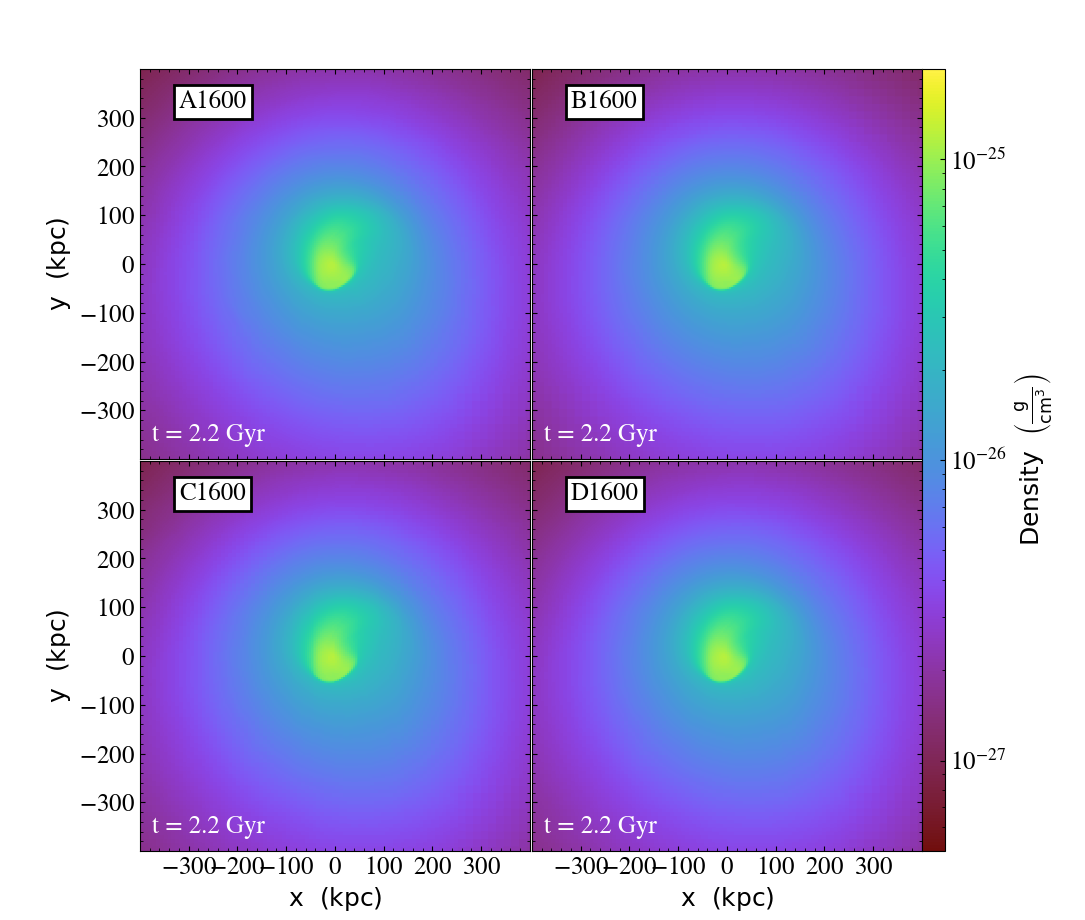}
 \caption{Density slice plots for high plasma $\beta$ 
 ($\beta \sim 1600$) simulations at $t=2.2$ Gyr. The simulations are still in the early phase of the cluster merger. At this epoch, the cold fronts are already beginning to form, though there are no significant differences among the four models.}
 \label{den_1600_2.2}
\end{figure}

\begin{figure}
 \includegraphics[width=\columnwidth]{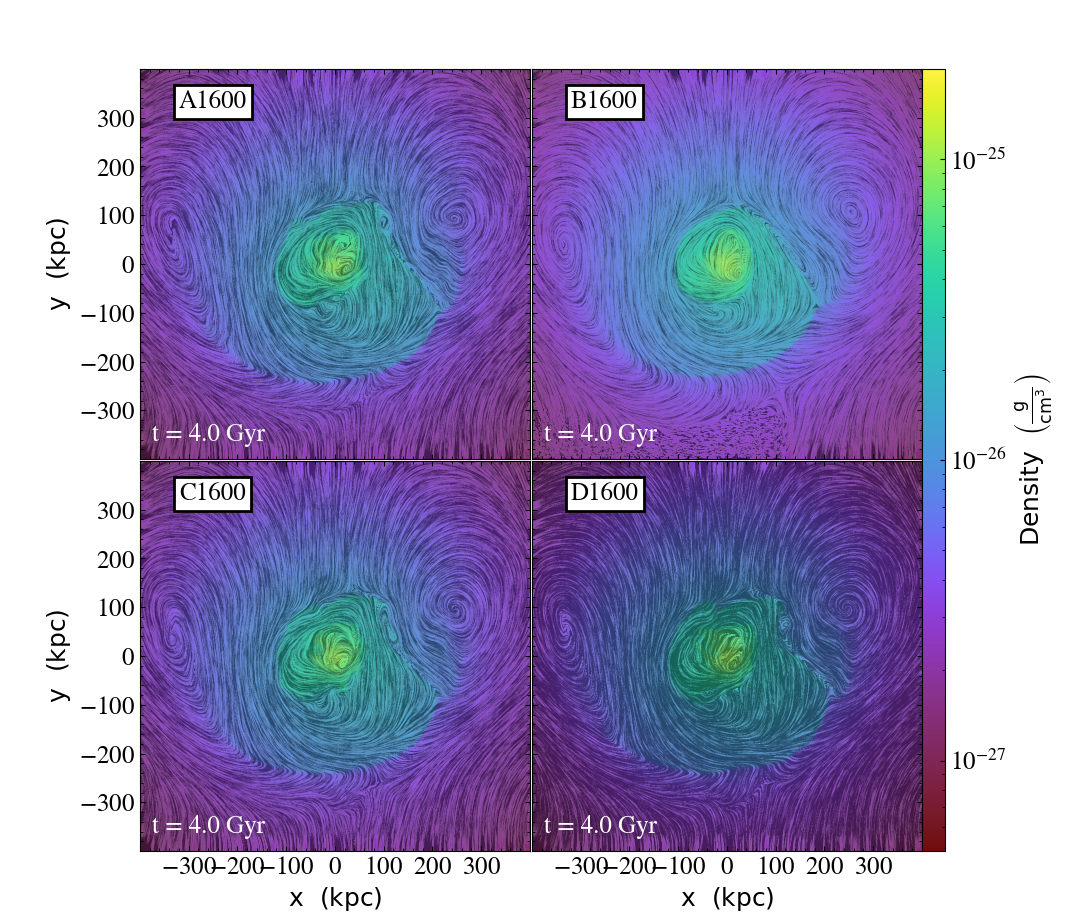}
 \caption{Density slice plots for high plasma $\beta$ ($\beta \sim 1600$) simulations at $t=4.0$ Gyr. Spiral-shaped cold fronts are apparent at this stage. The apparently stability of the cold front surfaces follows this order: $\rm{B}1600>\rm{C}1600>\rm{D}1600\gtrsim\rm{A}1600$. The black lines represent line integral convolution of the velocity field in the $x$–$y$ plane, used to visualise the gas velocity structure.} 
 \label{den_1600_4}
\end{figure}

\begin{figure*}
 \includegraphics[width=0.9\textwidth]{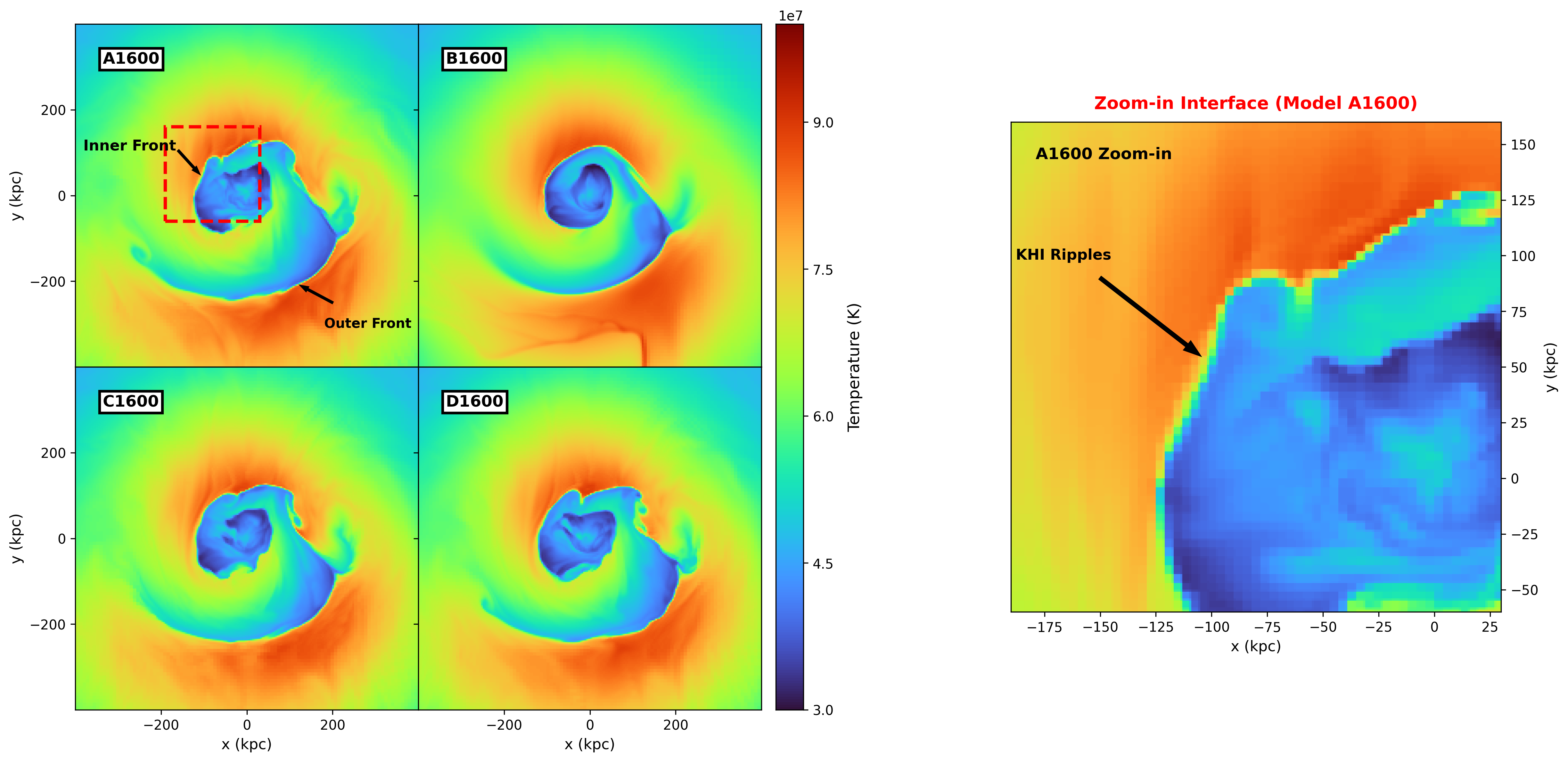}
 \caption{The left-hand panel is temperature slice plots for high plasma $\beta$ ($\beta \sim 1600$) simulations at $t=4.0$ Gyr. The black arrows indicate the locations of the cold fronts. Due to the sloshing motions induced by the merger, the cool-core region becomes disturbed, with hot outer gas intruding inward and forming temperature discontinuities. These structures, known as cold fronts, are characterized by the inner regions being cooler than the outer regions. The right-hand panel show zoomed-in views of the region enclosed by the red box in the A1600 temperature slice, highlighting ripples induced by the KHI.}
 \label{tem_1600}
\end{figure*}

Figure \ref{den_1600_2.2} shows slices of gas density at $t$ = 2.2 Gyr for cases A1600-D1600 when the cold fronts are just beginning to develop. Since the cold front structures are not yet well developed in the central region at $t$ = 2.2 Gyr, there are only minimal differences among the four cases. The structural similarity at this stage arises because there has not been sufficient time for the KHI to grow. In contrast, significant differences become apparent later ($t$ = 4.0 Gyr).

Figure \ref{den_1600_4} shows the density slices at $t$ = 4.0 Gyr. At this time, the cold front structures have become larger and more apparent. The KHI develop due to the velocity shear at the interface of sloshing cold fronts, which can be easily seen from the velocity streamlines overlaid. When the KHI develop, the cold fronts become unstable and small structures form along their interfaces such as the ripples in Figure \ref{den_1600_4}. The degree of the KHI varies between different viscosity models. In the later part of our results, the main focus will be on $t$ = 4.0 Gyr. Figure \ref{tem_1600} presents slices of the temperature for cases A1600-D1600 and zoom-in picture on the A1600 cold front surfaces. From the temperature slices one can more easily identify the cold fronts, as the gas temperature transitions from low to high values across the interfaces, as illustrated in Figure \ref{tem_1600}. The right panel of Figure \ref{tem_1600} shows a zoom-in on the cold front of the A1600 simulation, showing that the KHI manifests as ripples along the sloshing surface. Temperature plays an important role in our simulations because the Spitzer viscosity is a strong function of temperature, following Equation \ref{f7}. 

Case A1600 exhibits more ripples along the cold fronts compared to other cases due to the presence of KHI, which arises from the absence of viscosity. In contrast, the cold fronts in Case B1600 show very smooth surfaces. The presence of isotropic viscosity at the full Spitzer strength prevents the KHI from developing entirely. In Cases C1600 and D1600 involving anisotropic viscosity, the ripples are more pronounced than in the isotropic case at the cold front interfaces. This is because momentum transport by viscosity is limited to the local magnetic field direction, resulting in a weaker effective suppression of the KHI. In Case D1600, due to the limits on the pressure anisotropy imposed by the plasma microinstabilities, the viscous flux and therefore the suppression of KHI is further reduced, making Cases A1600 and D1600 very similar. These differences are most pronounced for the inner cold fronts within $r \sim 100$~kpc of the cluster center. As for the outermost cold front at a radius of $\sim$200 kpc, it has a very similar appearance in Cases A1600, C1600, and D1600, which can be attributed to the lower shear velocity across the surface that results in a longer timescale for the KHI to develop, and a higher viscosity due to the larger temperature (see Section \ref{subsec:3.2}).

\begin{figure}
 \includegraphics[width=\columnwidth]{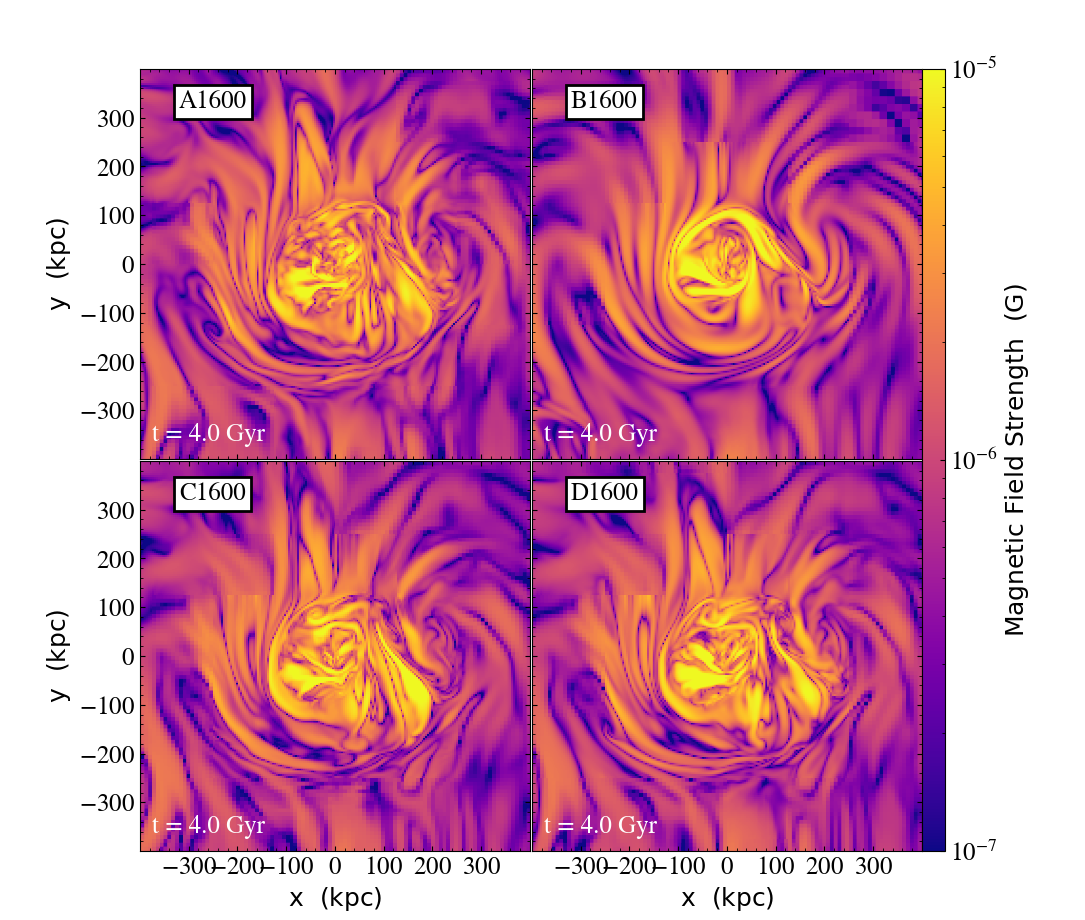}
 \caption{Magnetic field strength slice plots for high $\beta$ ($\beta \sim 1600$) simulations at $t=4.0$ Gyr. Velocity shear across the cold fronts stretches the magnetic field lines, leading to magnetic field amplification and the formation of magnetized layers aligned along the cold front surfaces.}
 \label{mag_1600}
\end{figure}

\begin{figure}
 \includegraphics[width=\columnwidth]{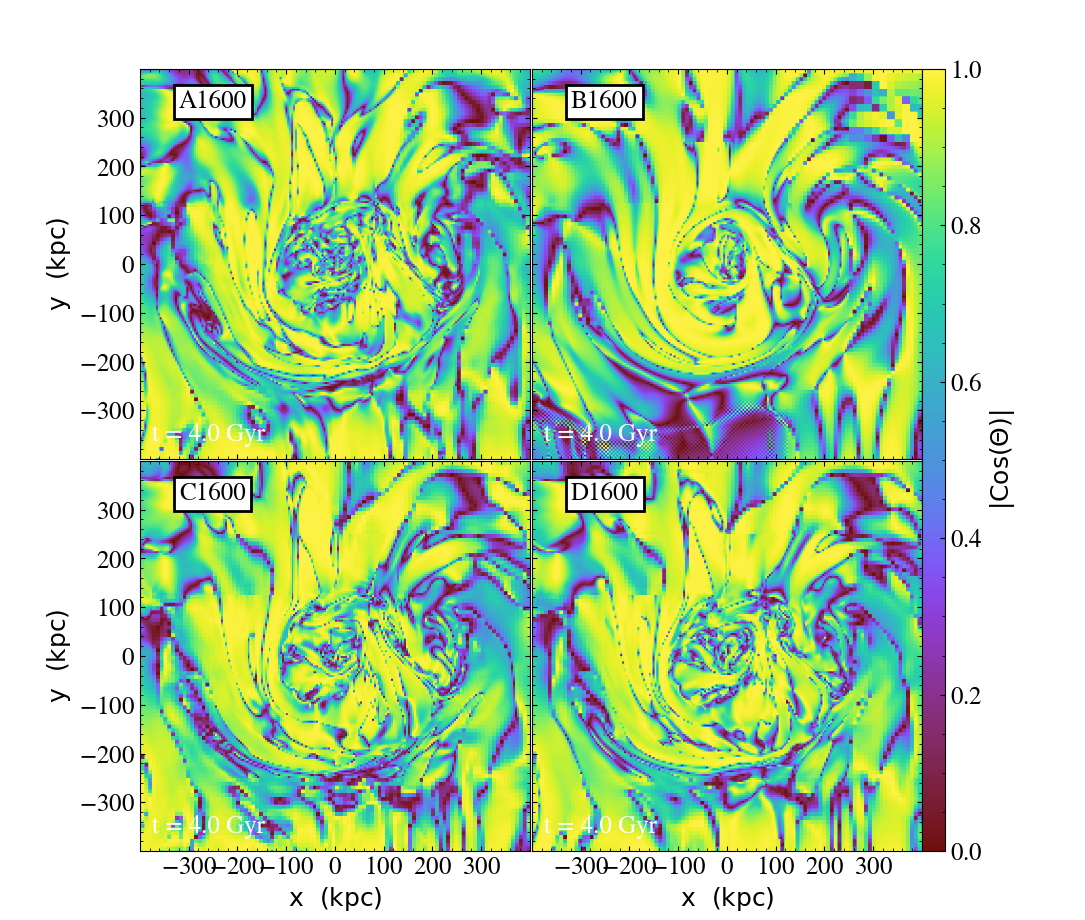}
 \caption{Angle between the magnetic field and velocity slice plots for high $\beta$ ($\beta \sim 1600$) simulations at $t=4.0$ Gyr. Because of the weaker magnetic tension in the high $\beta$ environment, the magnetic fields are more easily stretched by sloshing motions, enhancing their alignment (yellow regions).}
 \label{angle_1600}
\end{figure}

\begin{figure}
 \includegraphics[width=\columnwidth]{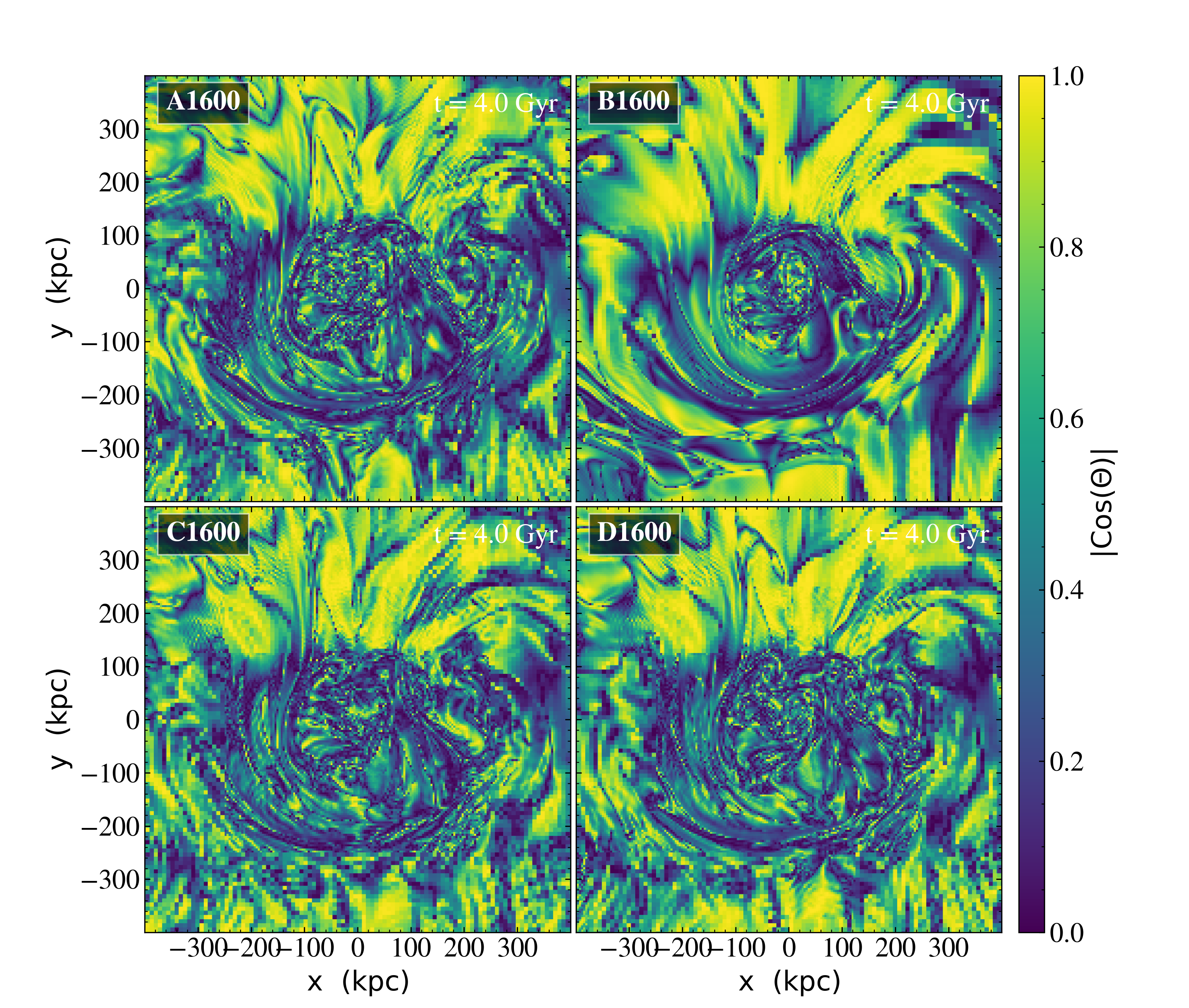}
 \caption{Angle between the magnetic field and temperature gradient slice plots for high $\beta$ ($\beta \sim 1600$) simulations at $t=4.0$ Gyr. Purple regions ($|\cos{\theta}| \sim 0$) indicate where the magnetic field is perpendicular to the temperature gradient, highlighting the magnetic draping effect. In these regions, magnetic field lines tend to align with the surfaces of the cold fronts.}
 \label{mag_draping_1600}
\end{figure}

Figure \ref{mag_1600} presents slices of magnetic field strength for Cases A1600-D1600.  During sloshing, the magnetic field lines are dragged along by the gas, causing the magnetic field to align with the surface of the cold fronts \citep{Lyutikov2006,Dursi2007,Keshet2010,ZuHone(2011)}. The velocity shear across the cold front stretches the magnetic field lines, thereby amplifying the magnetic field. As a result, the magnetic field becomes stronger and more ordered along the surface of the cold fronts. In Figure \ref{mag_1600}, the magnetic structures show the morphology of the layers at the interface of the cold fronts. The direction of these magnetic layers aligns with the surface of the cold fronts due to the sloshing motions. Figure \ref{angle_1600} presents slice plots of the angle $\theta$ between the magnetic field and velocity vectors for the $\beta \sim 1600$ cases, expressed in terms of $\cos{\theta}$. The yellow regions correspond to areas where $|\cos{\theta}| \approx 1$, indicating a strong alignment between the directions of the magnetic field and the velocity. The alignment just inside the cold front interfaces is more prominent than outside, where the higher velocities arise from sloshing motions. The morphological similarity between Figures \ref{mag_1600} and \ref{angle_1600} provides strong evidence for this alignment effect, supporting the interpretation that sloshing-induced velocity shear organizes the magnetic field into layers that follow the cold front surfaces. Furthermore, Figure \ref{mag_draping_1600} presents slice plots of the angle $\theta$ between the magnetic field and the temperature gradient as a key diagnostic for the magnetic draping effect \citep{Lyutikov2006, Dursi2007}. As the core sloshes, the ambient magnetic field is swept up and accumulates along the cold front surfaces. This process causes the magnetic field lines to become perpendicular to the local temperature gradient (the normal vector of the cold front surface). The magnetic draping effect, together with sloshing-induced shear motions, result in the aforementioned alignment effect, where the magnetic field forms layered structures along the cold front.

\subsection{Effects of viscosity in the $\beta$ = 1600 cases}
\label{subsec:3.2}



\begin{figure}
 \includegraphics[width=\columnwidth]{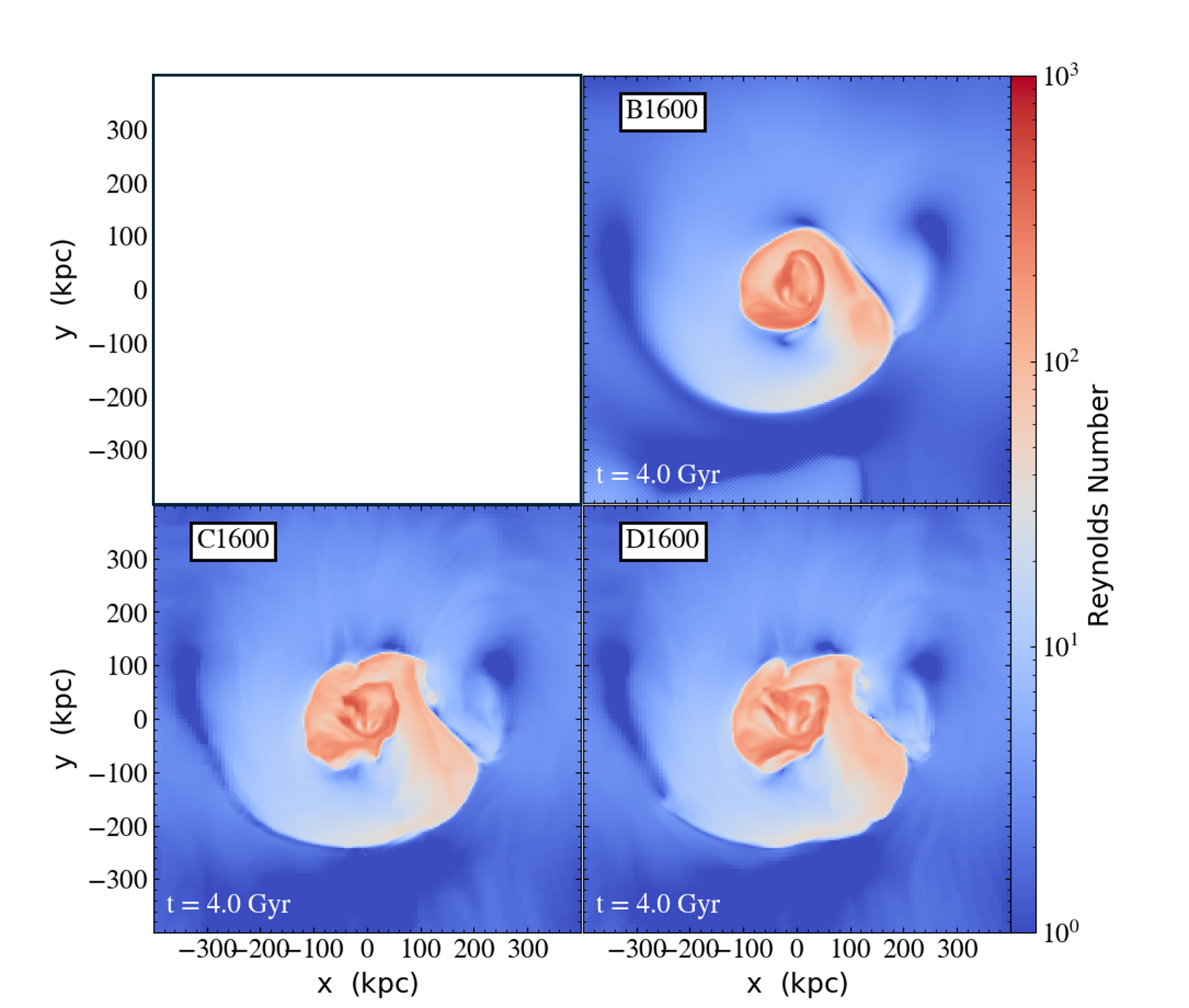}
 \caption{Reynolds number slice plots for high $\beta$ ($\beta \sim 1600$) simulations at $t=4.0$ Gyr. High Reynolds numbers in the central region indicate that the role of viscosity is less significant, favoring the development of the KHI and turbulence.}
 \label{rey_1600}
\end{figure}

\begin{figure}
 \includegraphics[width=\columnwidth]{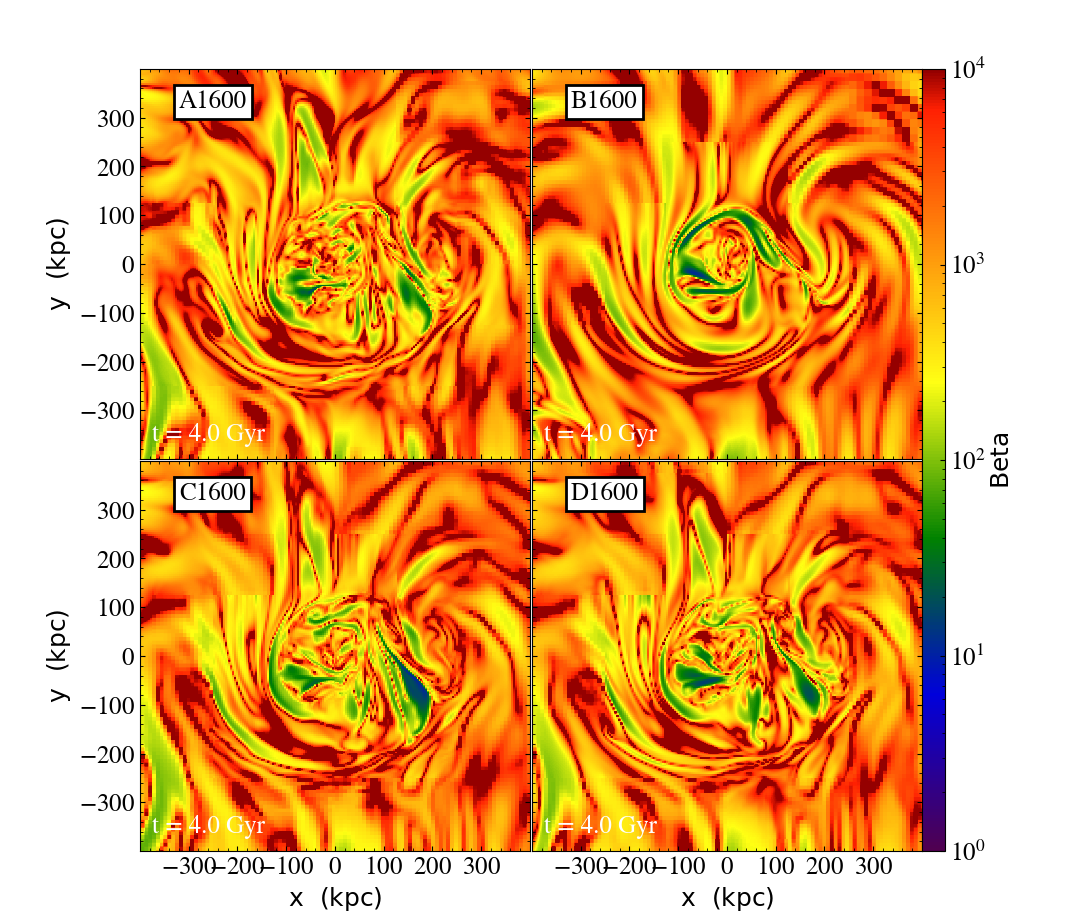}
 \caption{Slices of plasma $\beta$ at $t=4.0$ Gyr for Cases A1600-D1600 simulations (initial $\beta=1600$). The overall plasma $\beta$ values are high, indicating that even after shear amplification by the sloshing motions that magnetic fields are relatively weak and do not play a dominant role in the dynamics.}
 \label{beta_1600}
\end{figure}

 Figure \ref{rey_1600} shows the slices of Reynolds number for Cases B1600-D1600 at $t=4.0$ Gyr. The Reynolds number is defined as $Re=\rho vL/\mu$, where $v$ is fluid velocity, $L$ is characteristic length, and $\mu$ is the dynamic viscosity coefficient defined in Equation \ref{f7}. We choose $L = 30$ kpc as the characteristic length for the scale of the sloshing cold front. Case A1600 is excluded due to the absence of viscosity in this case. The Reynolds number provides a straightforward estimate of the strength of viscosity and where its effects become significant. For large Reynolds numbers, the inertial forces significantly exceed viscous forces, leading to turbulence and the KHI. Conversely, for smaller Reynolds numbers, viscous forces are more important, with opposite effects. The Reynolds number is typically defined assuming isotropic viscosity. In the anisotropic cases (Cases C1600 and D1600), the situation becomes more complex, making it difficult to directly compare the efficacy of viscosity among different models using the Reynolds number alone. Nevertheless, it remains a useful diagnostic in terms of where viscosity could play a role. \citet{Roediger2013a} demonstrated using plane-parallel simulations of the development of KHI at cold front-like surfaces that the critical Reynolds number on the hotter side of the cold front for the suppression of KHI is $\sim$30 (assuming a density contrast of $\sim$2 and isotropic viscosity with a Spitzer dependence). 
 
 One can see from Figure \ref{rey_1600} that the Reynolds numbers in the inner regions of the cold fronts are higher than those in the outer regions, primarily due to the temperature and velocity structure of the ICM. As shown in Figure \ref{tem_1600} and discussed in Section \ref{subsec:3.1}, the central part of the main cluster has lower temperatures, which leads to a smaller dynamic viscosity coefficient and a higher Reynolds number in the core region. Also, the gas velocities are higher within the cold front surfaces, also increasing the Reynolds number. Consequently, the inner regions are more prone to instability, making it easier for turbulence to develop through the growth of the KHI. In Figure \ref{rey_1600}, the dark blue regions just outside the cold front surfaces indicate that the effect of viscosity is stronger above the fronts, primarily due to the higher temperatures. In our simulation, the density contrast across the outer cold front (radius $\sim 200$ kpc) is $\sim 1.8$, matching the baseline in \citet{Roediger2013a}. Interestingly, the inner cold front (radius $\sim 100$ kpc) exhibits a higher density contrast of $\sim3.1$. This higher contrast has profound implications for viscous KHI suppression. For a Spitzer-like temperature-dependent viscosity, the ratio of Reynolds numbers between the cold and hot phases scales steeply with density contrast: $Re_{\rm cold}/Re_{\rm hot}=(\rho_{\rm c}/\rho_{\rm h})^{7/2}$, where $\rho_{\rm c}$ and $\rho_{\rm h}$ represent gas density on the cold- and hot-side of the front, respectively. Consequently, 
 even when the hot side falls below the critical threshold ($Re_{\rm hot} \lesssim 30$) and remains stable, the cold side will have a higher Reynolds number and can therefore develop small-scale structures. Because the two cold fronts are at different evolutionary stages, the hot-side Reynolds number is $\lesssim 30$ for the inner cold front and $\lesssim 1$ for the outer cold front. Since these values are at or below the critical Reynolds number proposed by \citet{Roediger2013a}, viscous effects in the hot layer are expected to effectively suppress the KHI. The smooth morphology of the cold fronts in Case B1600 is consistent with this expectation. Cases C1600 and D1600 have slightly higher Reynolds numbers in the hot layer compared to B1600. This is mainly due to the tangential velocity in B1600 being lower than in Cases C1600 and D1600 due to stronger effective viscosity. The slightly higher Reynolds number, together with additional suppression of viscosity due to anisotropic effects and plasma instabilities, gives rise to cold fronts that are less smooth, seen in Cases C1600 and D1600. 

 The magnetic field can also play a crucial role in shaping the cold fronts. Magnetic tension suppresses the KHI of the cold front when the direction of the magnetic field aligns with the cold front surface and when the magnetic tension is strong enough \citep{ll1960,Chandrasekhar1961}. 
 Additionally, the direction of the magnetic field can influence the suppression of KHI by viscosity, due to its anisotropic nature with respect to the local field line direction. In order to discern whether the KHI suppression is mainly caused by viscosity or magnetic tension, in Figure \ref{beta_1600} we present slices of plasma $\beta$ for Cases A1600-D1600. \citet{ZuHone(2011)} showed that the suppression of KHI by magnetic tension becomes significant when the plasma $\beta$ drops below $\sim 10$ (specifically, $\beta \lesssim (2/\gamma) M^{-2} \sim 5\text{--}13$ for $\gamma = 5/3$, derived from their Eq. 36 assuming characteristic subsonic shear flows across cold fronts with Mach numbers $M \sim 0.3\text{--}0.5$). Figure \ref{beta_1600} shows that the plasma $\beta$ near the cold front surfaces of all four cases ranges from $\sim$100 to 1000, suggesting that the suppression of the KHI in the high-$\beta$ cases is primarily due to viscosity rather than magnetic tension. Moreover, because of the initially high plasma $\beta$ values, the magnetic field lines in Cases A1600-D1600 are largely passively evolved and stretched by the sloshing motions. As shown in Figures \ref{mag_1600} and \ref{angle_1600} in Section \ref{subsec:3.1}, the magnetic field exhibits layered structures well aligned with the cold front surfaces. This magnetic field configuration further aids the suppression of the KHI by anisotropic viscosity in Cases C1600 and D1600. 

\begin{figure}
 \includegraphics[width=\columnwidth]{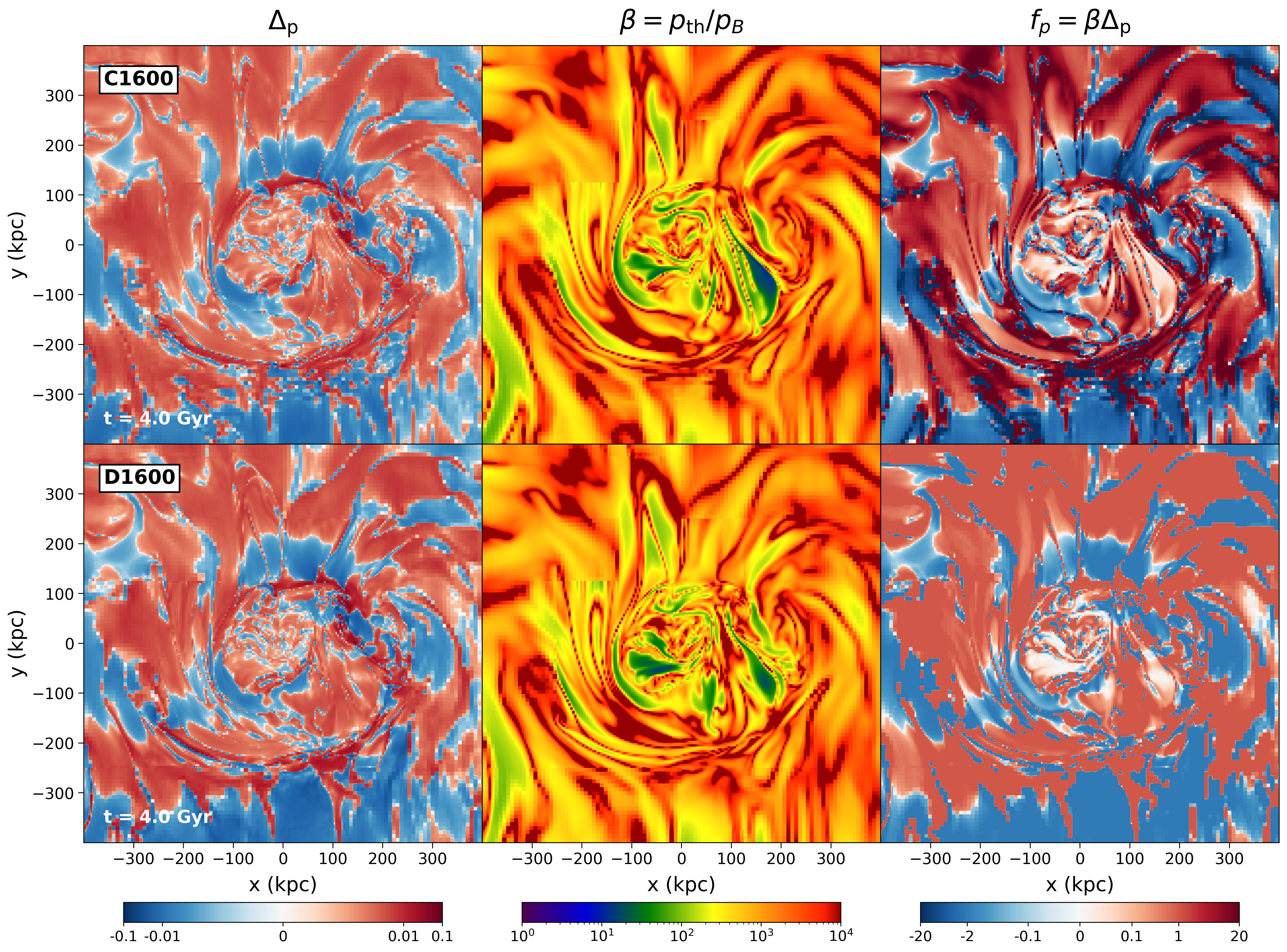}
 \caption{Slice plots illustrating the spatial distributions of the pressure anisotropy ($\Delta_{\rm p}$, left), plasma beta ($\beta = p_{\rm th}/p_{\rm B}$, middle), and the microinstability parameter ($f_{\rm p}=\beta\Delta_{\rm p}$, right) at $t=4.0$ Gyr for the high-$\beta$ ($\beta=1600$) simulations. The top row shows Case C1600 (anisotropic viscosity), while the bottom row presents Case D1600 (anisotropic viscosity with pressure-limiters). Positive values of $f_{\rm p}$ correspond to the mirror-instability regime, whereas negative values correspond to the firehose-instability regime. Owing to the high overall plasma $\beta$, large positive and negative values of $f_{\rm p}$ develop over broad regions, particularly around the sloshing cold fronts, indicating that the local pressure anisotropy frequently exceeds the microinstability thresholds.}
 \label{H_2x3}
\end{figure}

Figure \ref{H_2x3} presents slice plots of the pressure anisotropy magnitude ($\Delta_{\rm p}$, left), plasma beta ($\beta = p_{\rm th}/p_{\rm B}$, middle), and the microinstability parameter ($f_{\rm p} = \beta \Delta_{\rm p}$, right) that measures the degree of departure from the microinstability threshold. For case D1600, due to microinstability limiter is imposed, $f_{\rm p}$ be restrict between -2 and 1. 
As shown in the figure, due to the high background plasma $\beta$, the unrestricted $f_{\rm p}$ naturally reaches very large values across most of the domain, particularly within the active sloshing region. It is also worth noting that the sloshing cold fronts tend to exhibit $f_{\rm p} \gtrsim 1$ (e.g., the upper-right corner of the figure), indicating that these regions are susceptible to the mirror instability. This is a consequence of the amplification of the magnetic field by velocity shear and compression associated with the sloshing motions. We calculate the volume fraction within the inner core region ($r < 400$ kpc) where the plasma satisfies the microinstability criterion ($f_{\rm p} < -2$ or $f_{\rm p} > 1$). We find that approximately 59\% of the volume in Case C1600 lies outside the bounds. Because the pressure anisotropy is bounded by the microinstability limit across this large volume, the effective anisotropic viscosity in Case D1600 is reduced. This explains the morphological differences observed between Cases C1600 and D1600.

The simulation results are consistent with expectations. Case A1600, which lacks viscosity, exhibits the most unstable cold fronts. In contrast, Case B1600, which includes isotropic viscosity, produces the smoothest and most stable cold fronts. Both Case C1600 and Case D1600 employ anisotropic viscosity, but produce different outcomes. Case C1600 maintains relatively stable cold fronts, while Case D1600, in which the viscosity is limited by pressure-anisotropy bounds associated with microinstabilities, exhibits small ripple-like structures along the cold front surfaces similar to the inviscid case.

\subsection{Simulations with $\beta=100$}
\label{subsec:3.3}

\begin{figure}
 \includegraphics[width=\columnwidth]{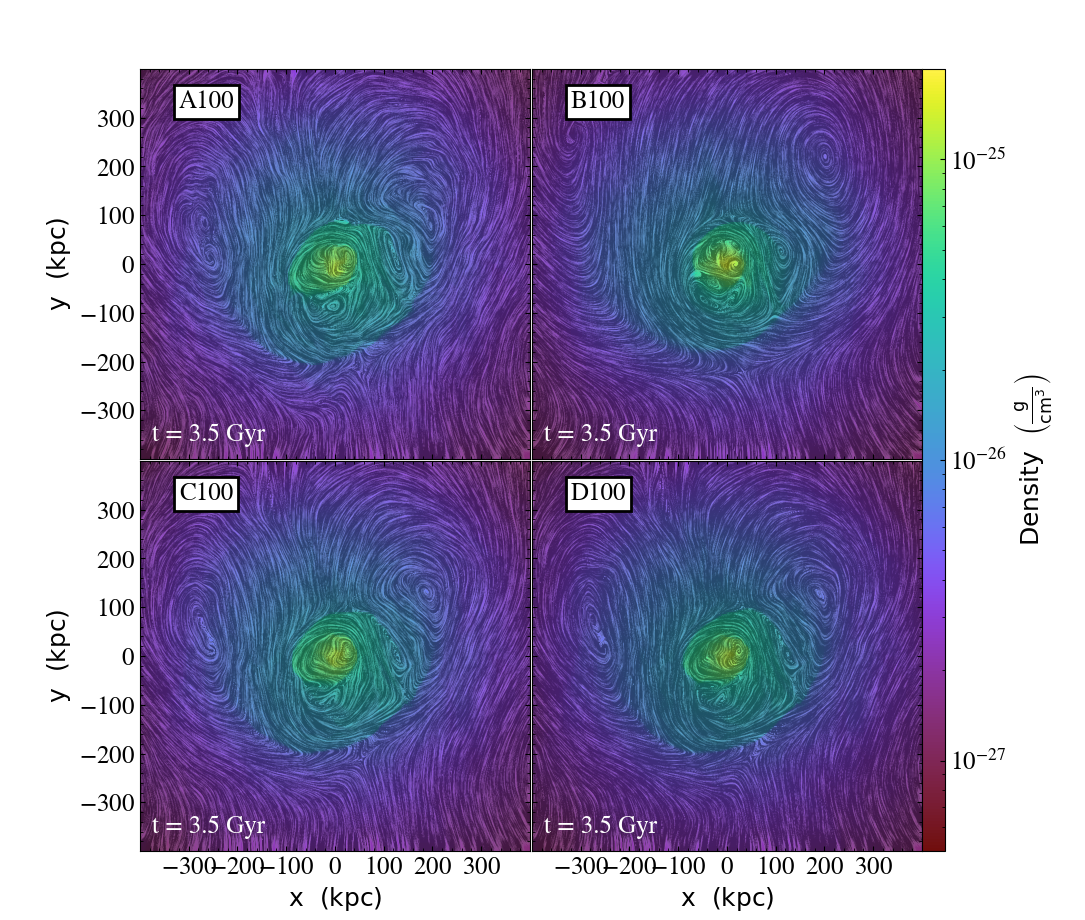}
 \caption{Density slice plots for the $\beta \sim 100$ simulations at $t=4.0$ Gyr. With a stronger initial magnetic field, all four models exhibit more stable cold front surfaces compared to the weaker field cases. The black lines represent line integral convolution of the velocity field in the $x$–$y$ plane, used to visualise the gas velocity structure.}
 \label{den_100}
\end{figure}

\begin{figure}
 \includegraphics[width=\columnwidth]{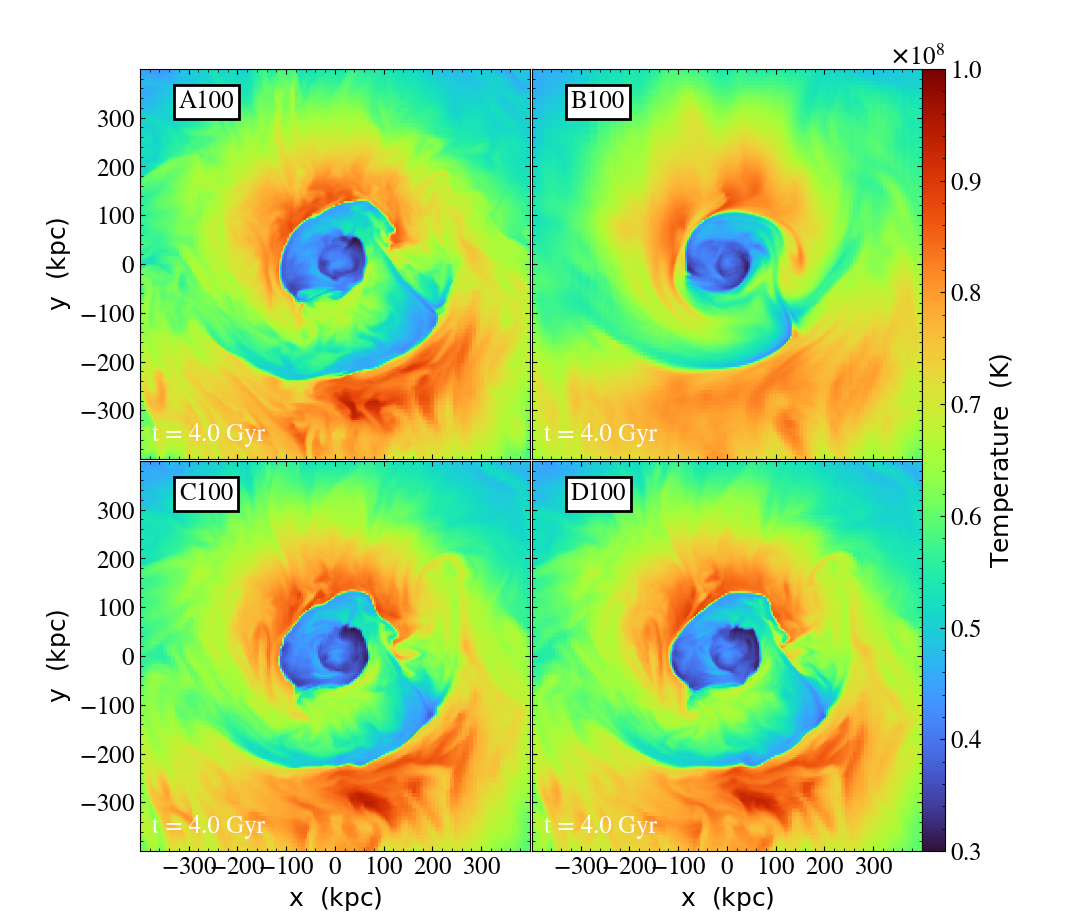}
 \caption{Temperature slice plots for the $\beta \sim 100$ simulations at $t=4.0$ Gyr, highlighting the same features as Figure \ref{den_100}.}
 \label{tem_100}
\end{figure}

\begin{figure}
 \includegraphics[width=\columnwidth]{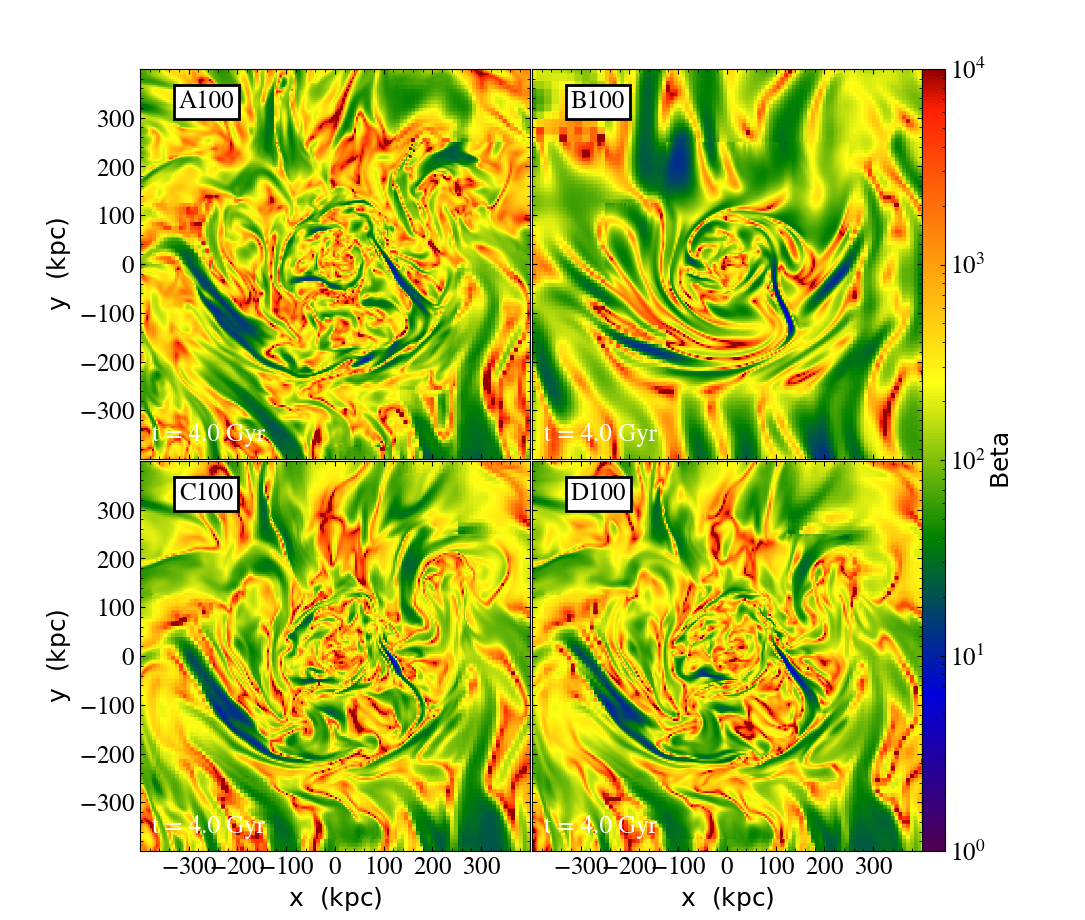}
 \caption{Slice plots for the plasma $\beta$ at $t=4.0$ Gyr for the $\beta \sim 100$ simulations. At some of the cold front surfaces, $\beta \sim $ 10 or even lower.}
 \label{beta_100}
\end{figure}

\begin{figure}
 \includegraphics[width=\columnwidth]{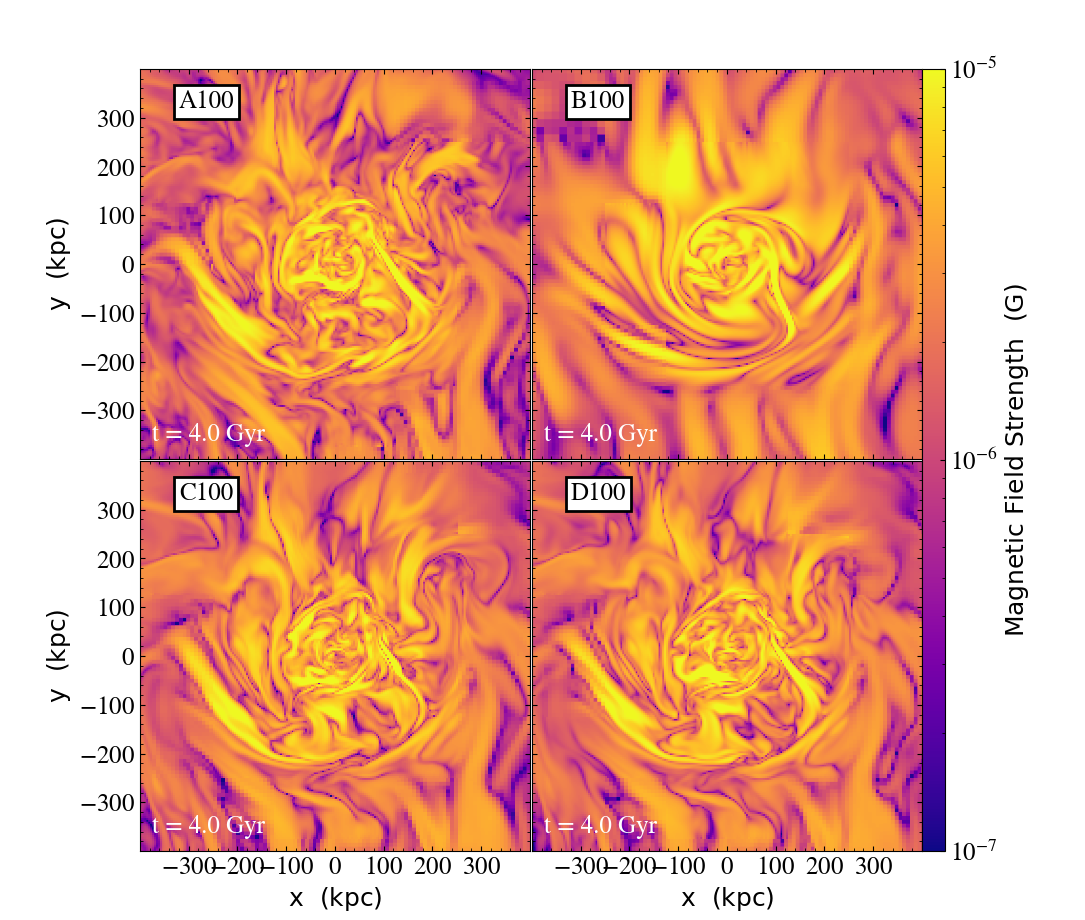}
 \caption{Magnetic field strength slice plots for low $\beta$ ($\beta \sim 100$) simulations at $t=4.0$ Gyr. Although the magnetic fields form layer-like structures similar to those in the $\beta=1600$ cases, its overall structure is more turbulent and disordered.}
 \label{mag_100}
\end{figure}

In this section, we present the low-$\beta$ cases to examine how the gas in the central region will evolve with a stronger initial magnetic field in addition to viscosity. Figures~\ref{den_100} and \ref{tem_100} show density and temperature slice plots, respectively, for the low-$\beta$ cases. As in Figure \ref{den_1600_4}, the line integral convolution shows the velocity streamlines, indicating the shears at the cold front surfaces. Due to the stronger magnetic fields, Cases A100, C100, and D100 exhibit smoother cold front surfaces compared to the high-$\beta$ = 1600 cases due to the suppression of the KHI by magnetic tension, in addition to the effect of viscosity where it is present. This is supported by Figures \ref{beta_100} and \ref{mag_100}, which shows the plasma $\beta$ and magnetic field strength slice plots at $t = 4.0$ Gyr for the low-$\beta$ cases. For certain regions along the cold front surfaces, the $\beta$ values drop to $\sim$10 or even lower because of the local enhancement of magnetic field strength due to stretching by the sloshing motions, consistent with the results of previous work by \citet{ZuHone(2011)}. Unlike cases C1600 and D1600, cases C100 and D100 are very similar in appearance, due to the wider pressure anisotropy bounds (Eq.\ \ref{f9}) when the plasma $\beta$ is lower, which thus permit a larger viscous suppression than the $\beta = 1600$ simulations. As in the $\beta = 1600$ simulations, the inner cold fronts within $r \sim 100$~kpc exhibit more indications of KHI (and thus more differences between the different cases with different viscosity models) than the outer cold front at $r \sim 200$~kpc.

\begin{figure}
 \includegraphics[width=\columnwidth]{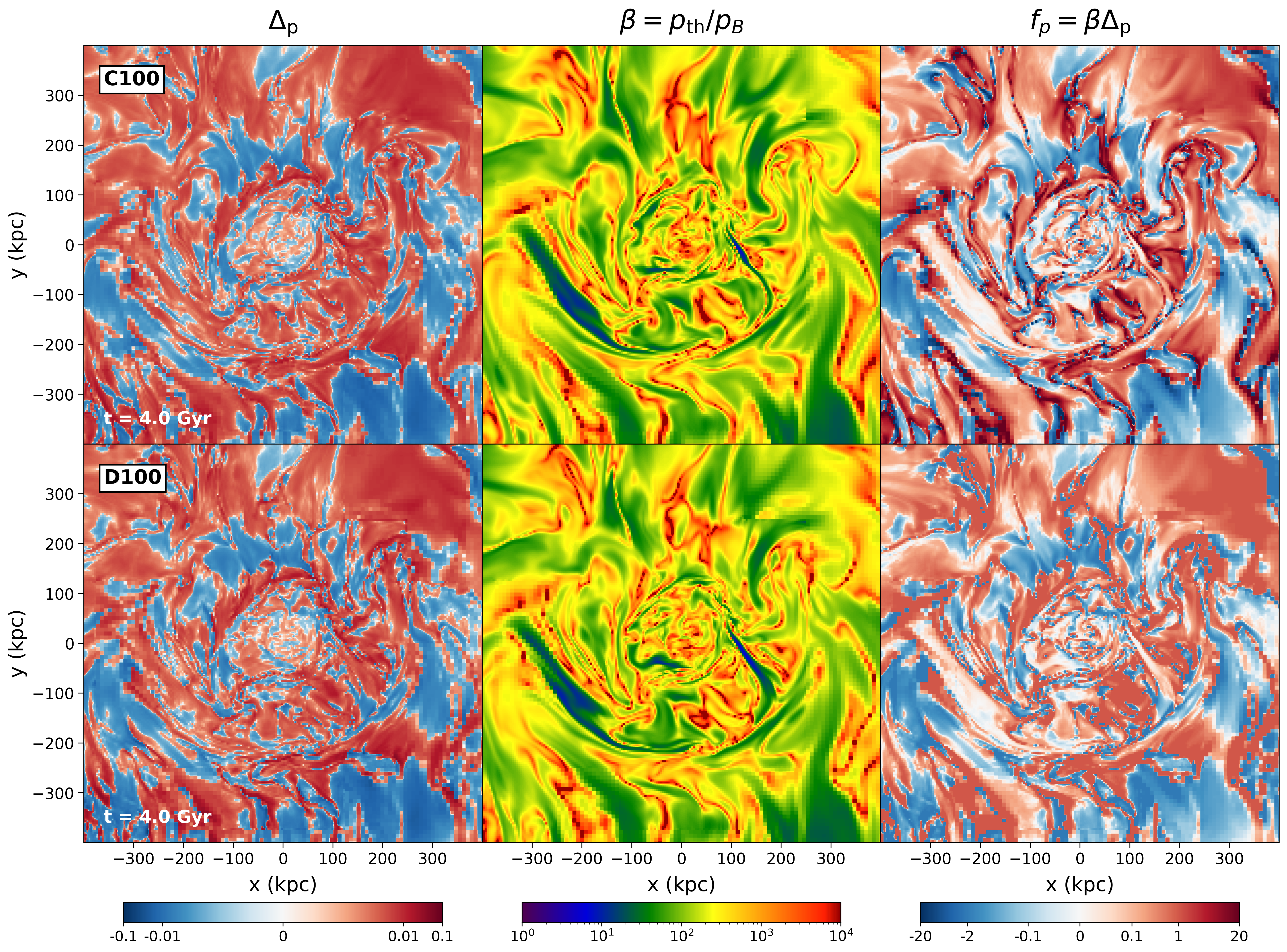}
 \caption{Same as Figure \ref{H_2x3}, but for the low-$\beta$ ($\beta = 100$) simulations. In contrast to the widespread instability seen in the high-$\beta$ cases, the stronger background magnetic field here restricts the regions where the microinstability thresholds ($f_{\rm p}>1$ or $f_{\rm p}<-2$) are exceeded. Consequently, regions exceeding the microinstability thresholds are primarily confined to the highly sheared cold-front boundaries.}
 \label{L_2x3}
\end{figure}

Figure \ref{L_2x3} presents the same microinstability diagnostics as Figure \ref{H_2x3}, but for Cases C100 and D100. While the overall magnitude of the pressure anisotropy ($\Delta_{\rm p}$, left panel) remains comparable to that in the high-$\beta$ cases, the significantly stronger background magnetic field (and thus lower $\beta$) substantially reduces the volume where the microinstability thresholds ($f_{\rm p} \gtrsim 1$ or $f_{\rm p} \lesssim -2$) are exceeded. Consequently, regions susceptible to the mirror ($f_{\rm p} > 1$) and firehose ($f_{\rm p} < -2$) instabilities become much less common. The volume fraction within the inner core region ($r<400$ kpc) for which the plasma satisfies the instability thresholds decreases to approximately 21\% in Case C100. Thus Case D100 retains a similar viscous flux to C100, making Cases C100 and D100 very similar in appearance. 

\begin{figure}
 \includegraphics[width=\columnwidth]{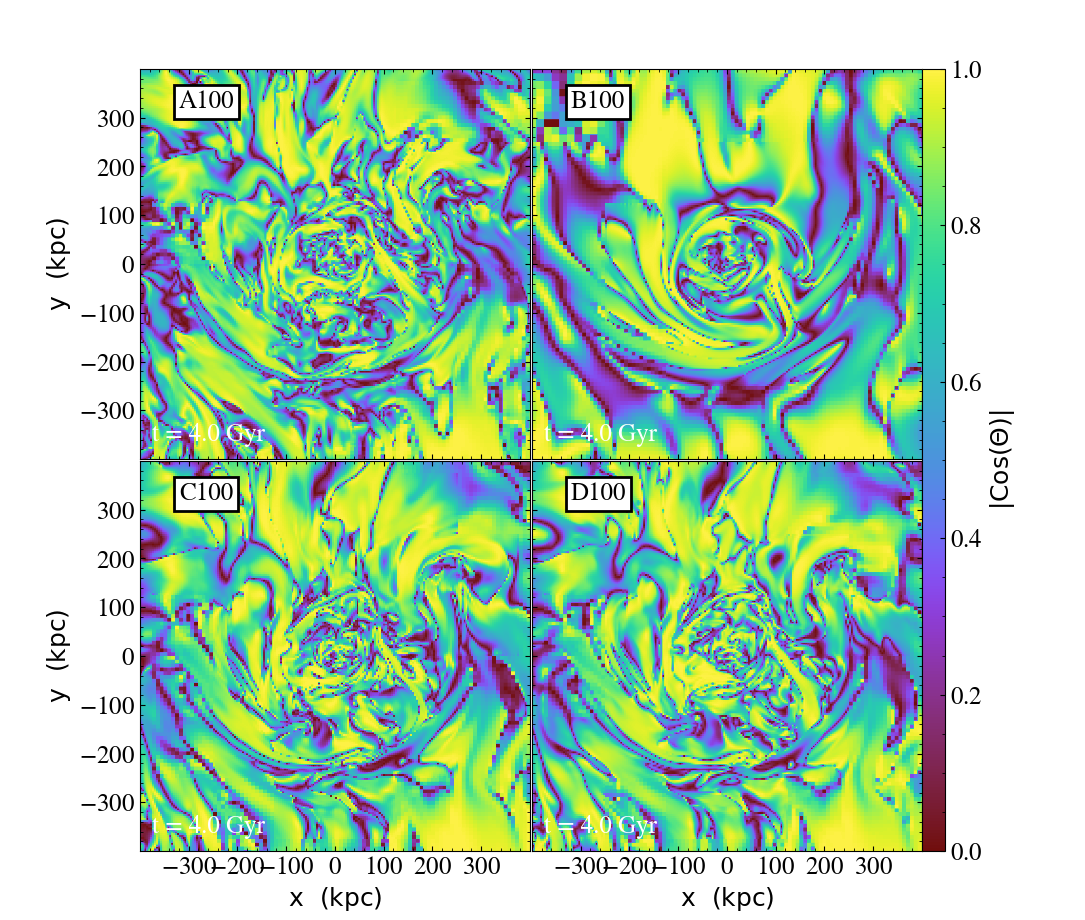}
 \caption{Angle between magnetic field and velocity slice plots for low $\beta$ ($\beta \sim 100$) simulations at $t=4.0$ Gyr. We use $|\cos{\theta}|$ to present the figure, which ranges from 0 (not aligned) to 1 (aligned). There is a wider range of angles both inside and outside the cold fronts in these simulations than the $\beta$ = 1600 cases.}
 \label{angle_100}
\end{figure}

\begin{figure*}
 \includegraphics[width=0.9\textwidth]{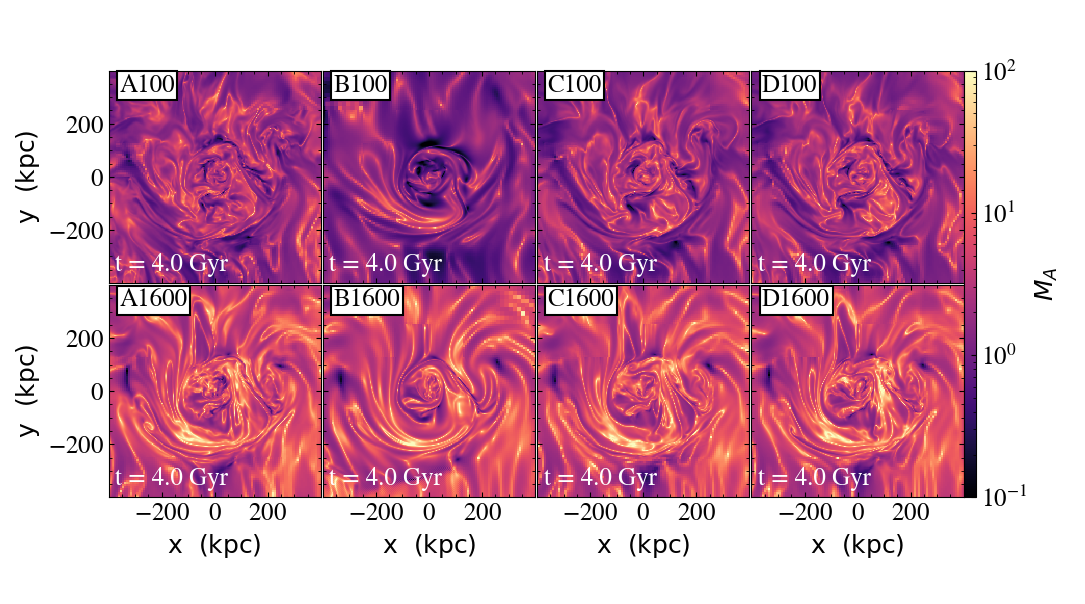}
 \caption{Alfvénic Mach number for all simulations at $t=4.0$ Gyr. The upper four slices correspond to the low-$\beta$ cases, while the lower four slices show the high-$\beta$ cases. The low-$\beta$ cases exhibit systematically lower Alfvén Mach numbers than the high-$\beta$ cases.}
 \label{Mach_all}
\end{figure*}


As we did for the $\beta = 1600$ simulations, we plot the angle between the magnetic field and the velocity vectors in Figure \ref{angle_100}. As in Figure \ref{angle_1600}, the yellow regions indicate areas where $|\cos{\theta}|$ is close to 1, signifying alignment between the magnetic field and the velocity directions produced by gas motions stretching the magnetic field. 
Compared to Figure \ref{angle_1600}, fewer regions have $|\cos{\theta}| \approx 1$, even near the cold front surfaces. In Cases A100, C100, and D100, these different degrees of alignment have a turbulent structure, as does the magnetic field strength itself. In Case B100, the isotropic viscosity damps turbulent motions such that the magnetic field structure is highly ordered, but there are nevertheless still clear divisions between regions where the magnetic field and velocity vectors are parallel and those where they are not. 

In the low-$\beta$ simulations, the magnetic tension is stronger, which results in the magnetic field lines becoming more resistant to becoming stretched by the velocity field. The most straightforward way to illustrate this is by the Alfv\'enic Mach number $M_A = v/v_A$, where $v_A = B/\sqrt{4\pi\rho}$ is the Alfvén speed. Figure \ref{Mach_all} shows the spatial distribution of $M_A$ for all of the simulations with both values of $\beta$. The contrast between the two cases is substantial: the high-$\beta$ simulations exhibit large Alfvénic Mach numbers (with most regions having $M_A > 5$), whereas the low-$\beta$ cases predominantly show $M_A < 2$. Large Alfvénic Mach numbers indicate that the plasma can more easily drag or bend the magnetic field lines, indicating that the magnetic field amplification and the alignment effects discussed above are more readily triggered. In contrast, in the low-$\beta$ cases, with a lower $M_A$, the magnetic fields are less easily transported and stretched. Though the resulting misalignment with the cold front surfaces in the $\beta = 100$ simulations would seem to lead to a lower suppression of KHI by the magnetic fields, the results are not that straightforward since most of the time there is some component of the field parallel to the front, and the overall field strength is higher than in the $\beta$ = 1600 cases. Nevertheless, magnetic draping still robustly occurs in the low-$\beta$ cases. While the stiffer magnetic field lines are harder to stretch (as indicated by the low $M_A$), the velocity of the sloshing cold fronts is still sufficient to sweep them up. Because the strong magnetic tension suppresses small-scale turbulent tangling, the resulting draped layer is highly coherent, making the draping signature ($|\cos \theta| \sim 0$) even more visually pronounced than in the high-$\beta$ cases. 


\section{Discussion}
\label{sec:Discussion}

Our research follows the setup and discussion outlined by \cite{ZuHone(2011),ZuHone_2015}. In \cite{ZuHone(2011)}, the authors investigated how varying initial magnetic field strengths affect the sloshing motion in galaxy clusters. In a follow-up study, \cite{ZuHone_2015} explored the impact of different viscosity models on the evolution and stability of cold fronts, though with only a weak initial magnetic field with $\beta = 1000$, and without considering the effect of microinstabilities. Here, we try to combine those two works and aim to delve deeper into the influence of magnetic field, and the case anisotropy viscosity limited by microinstabilities.

In the high-$\beta$ simulation ($\beta \sim 1600$), our results are consistent with those of \cite{ZuHone_2015}. Case A fails to suppress the KHI, while Cases B (isotropic viscosity) and C (anisotropic viscosity) demonstrate varying degrees of KHI suppression. Our study shows that the pressure-anisotropy suppression effect of the microinstabilities (case D) further limits the ability of viscosity to suppress the KHI. Our results are also consistent with \citet{ZuHone(2011)}, in that a stronger initial magnetic field provides stronger suppression of KHI. The stronger magnetic field also permits the pressure anisotropy to grow larger in case D100, which results in stronger viscous suppression compared with case D1600. However, in the cases with high plasma $\beta$, the magnetic field is more easily stretched by the sloshing motions and aligned with the cold front surfaces, whereas in our low-$\beta$ cases the alignment effect is weaker. On the one hand, a magnetic field aligned with the front surface (as in our $\beta = 1600$ simulations) provides maximal KHI suppression for a given field strength, but a partially aligned field with stronger field strength overall (the $\beta = 100$ simulations) may nevertheless have a more noticeable effect. The effect on Braginskii viscosity from the varying magnetic field strength is twofold. The first is that a magnetic field more aligned with the front surface ($\beta = 1600$ cases) will prevent viscous suppression until KHI produces an alignment between the local magnetic field and velocity gradient, whereas if the field is more tangled ($\beta = 100$ cases) and is at least partially aligned with the velocity gradients to begin with, the suppression will be more immediate. Second, as already noted the stronger magnetic field permits stronger viscous flux. In complex situations such as modeled by our simulations, these effects can be difficult to disentangle from each other in a straightforward way. 

There is another effect of the stronger initial magnetic field in the low-$\beta$ simulations that can complicate the interpretation of our results. Simulations of idealized cluster mergers such as the ones we use in this work are initialized with each cluster in hydrostatic equilibrium, with a high-$\beta$ turbulent magnetic field added in. As shown by \cite{Chadayammuri_2022}, such a tangled field produces seed velocity fluctuations when the magnetic tension inherent in the initial field relaxes. The resulting velocity perturbations can serve as the seed fluctuations for the KHI, potentially influencing the cold front morphology. This effect is more pronounced for lower initial $\beta$. The result is that the same simulations which are expected to be most effective at suppressing KHI growth via magnetic tension are also the ones which will have the strongest initial velocity perturbations to seed the KHI. To quantify this effect, we calculated the velocity standard deviation at an early time ($t=1.2$ Gyr). The velocity standard deviation is $\sim 16$ km/s for B100 and $\sim 5$ km/s for B1600 within $<200$ kpc. This result is consistent with the expectation that greater initial field strength generates stronger velocity fluctuations that can seed the KHI. Nevertheless, although the fluctuations are larger in B100, their amplitudes remain small in both simulations, and we do not observe significant KHI growth at later times. This is consistent with magnetic tension effectively suppressing the instability.

Finally, in theory Cases D100 and D1600 should more closely represent realistic ICM conditions, as they incorporate the regulation of Braginskii viscosity by kinetic microphysics. While these cases do develop indications of KHI at cold fronts, such structures are not necessarily inconsistent with observations, since some cold fronts have been found to exhibit evidence of KHI in deep X-ray exposures \citep{Wang_2018,Werner_2016,Su_2017}. 

While this work was nearing completion, a similar study was being conducted by \cite{zuhone2026}, which investigates two relatively low-$\beta$ cases with Braginskii viscosity limited by microinstabilities. Our work was carried out independently and yields broadly consistent results, although some differences in the details remain due to variations in numerical resolution and initial conditions. In particular, the finest cell size in the other study is $\Delta x = 0.98$ kpc, which is four times smaller than in our simulations. This higher resolution allows the development of finer-scale structures along the cold fronts, as the KHI has higher growth rates for small-scale perturbations. The implementation of the microinstability constraint in their study also differs slightly from ours, as they have considered a few variations of the hard-wall limiters motivated by recent plasma simulations. In addition, we include the isotropic viscosity cases to provide a direct comparison with the other models, though they do not. Due to their higher resolution, the inviscid cases in their simulations exhibit finer small-scale structures 
than the microinstability-limited cases. Nevertheless, our results and conclusions are broadly consistent, particularly in that the inclusion of pressure-anisotropy bounds makes the suppression of KHI by viscosity less effective, resulting in a morphology that more closely resembles the inviscid case. Also, in both works, for low-$\beta$ cases, higher magnetic field strengths play an important role in suppressing the KHI and widening the possible range of pressure anisotropy when the effects of microinstabilities are included. 


\section{Conclusions}
\label{sec:Conclusions}
In X-ray observations of the ICM, we find cold fronts in many merging clusters, which are believed to arise from sub-sonic motions of low entropy gas. Some observed cold fronts appear smooth, while others exhibit disruptions which resemble the effects of KHI. This may indicate that while some degree of viscosity is indeed present in the ICM, but the degree of KHI suppression varies from system to system. Such an observational diversity provides a valuable opportunity to constrain the microphysical properties of the ICM through comparison with numerical simulations. In our study, we utilized simulations with varying ICM viscosity models to assess the suppression of the KHI on the surfaces of cold fronts, with the aim of exploring the roles of ICM viscosity and magnetic tension. Specifically, we explore the effects of viscosity on sloshing cold fronts in galaxy clusters, utilizing Braginskii-MHD simulations. Our research follows the methodologies established by \cite{ZuHone(2011)}, \cite{ZuHone_2015}, and \cite{Kingsland(2019)}, with the goal of delving deeper into the influences of magnetic fields and anisotropic viscosity limited by microinstabilities. We performed a series of cluster merger simulations including four different viscosity models (A: No viscosity; B: Isotropic Spitzer viscosity; C: Anisotropic Braginskii viscosity; D: Anisotropic viscosity limited by microinstabilities) and two values of plasma $\beta$.

Our simulation results in the high plasma $\beta$ cases ($\beta \sim$  1600) are summarized as follows:
\begin{itemize}
    \item Case A1600: Fails to suppress KHI, leading to disrupted cold fronts.
    \item Case B1600: Effectively suppresses KHI, resulting in smooth cold front surfaces.
    \item Case C1600: Demonstrates intermediate suppression of KHI, and the surfaces of cold fronts appear more stable.
    \item Case D1600: Does not effectively suppress KHI, similar to Case A, and forms small-scale perturbations at cold front surfaces.
\end{itemize}

These results suggest that, when the effect of magnetic tension is negligible, the smoothness of the cold fronts could in principle allow us to constrain the viscosity of the ICM. Several observational studies \citep{Wang_2018, Werner_2016, Su_2017} have reported ripple-like structures along cold front surfaces and derived upper limits on the effective viscosity, suggesting that the full isotropic Spitzer viscosity (as in Case B) is too strong (which is already largely assumed on theoretical grounds alone). However, the differences between the other viscosity models are more subtle, and are at least qualitatively compatible with existing observations. 


The simulation results for the low plasma $\beta$ cases ($\beta \sim$ 100) are as follows:
\begin{itemize}
    \item Case A100: The magnetic field suppresses the KHI, yielding smoother cold front surfaces compared to Case A1600.
    \item Case B100: Effectively suppresses the KHI, resulting in smooth cold front surfaces.
    \item Case C100: The magnetic field provides more suppression of KHI, resulting in a smoother cold front compared to Case C1600.
    \item Case D100: The suppression of the KHI is more effective compared to Case D1600 and similar to Case C100, due to the wider pressure anisotropy bounds.
\end{itemize}

In the simulations with low plasma $\beta$, the strong magnetic field suppresses the perturbations and hence it is more difficult for the KHI to grow compared to the high-$\beta$ simulations. Therefore, in the low-$\beta$ cases, the cold front morphology for all viscous models is more similar (though Case B100 still has the most stable surface). 


From the time when cold fronts were discovered shortly after the launch of \textit{Chandra}, there have been attempts to place constraints on the ICM viscosity or the strength of the magnetic field layers at cold front surfaces from the presence or absence of KHI. Our simulations suggest that both of these effects can play an important role in stabilizing cold fronts, but that their effects are difficult to disentangle from one another if the magnetic field is strong enough such that it both suppresses KHI by itself while also permitting a wider range of pressure anisotropies that will have the same effect.





\section*{Acknowledgements}

The authors thank the anonymous referee for their valuable comments and suggestions, which greatly improved the quality of this manuscript. The simulations were performed using computational resources provided by the Center for Informatics and Computation in Astronomy (CICA) at National Tsing Hua University, funded by the Ministry of Education (MOE) of Taiwan. MHH and HYKY acknowledge support from National Science and Technology Council (NSTC) of Taiwan (NSTC 112-2628-M-007-003-MY3; NSTC 114-2112-M-007-032-MY3). HYKY acknowledges support from Yushan Scholar Program of the Ministry of Education (MoE) of Taiwan (MOE-108-YSFMS-0002-003-P1; MOE-114-YSFMS-
0002-002-P2). Support for JAZ was provided by the {\it Chandra} X-ray Observatory Center, which is operated by the Smithsonian Astrophysical Observatory for and on behalf of NASA under contract NAS8-03060. FLASH was developed in part by the DOE NASA- and DOE Office of Science-supported Flash Center for Computational Science at the University of Chicago and the University of Rochester. Data analysis presented in this paper was conducted with the publicly available yt visualization software \citep{Turk_2011}.

\section*{Data Availability}

The data underlying this article will be shared upon
reasonable request to the corresponding author.



\bibliographystyle{mnras}
\bibliography{cite} 




\appendix

\section{Further analysis related to anisotropy viscosity}

In this Appendix, we present additional diagnostics associated with anisotropic viscosity to further elucidate the physical mechanisms responsible for suppressing the Kelvin--Helmholtz instability (KHI).

\begin{figure}
 \includegraphics[width=\columnwidth]{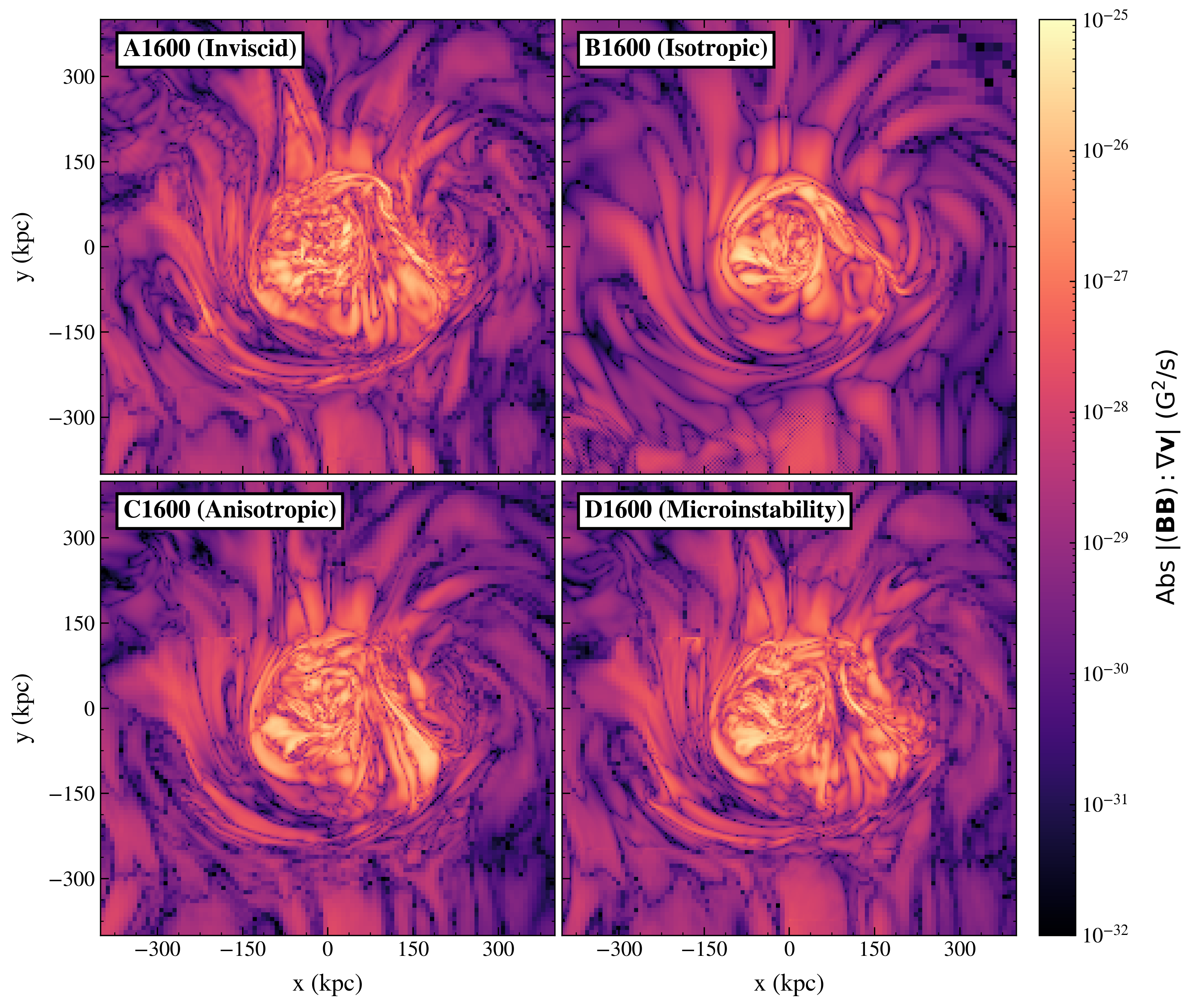}
 \caption{Slice plots displaying the magnitude of the magnetic field stretching term, $|(\vb{B}\vb{B}):\nabla\bm{v}|$, for the four high-$\beta$ ($\beta = 1600$) models at $t = 4.0$ Gyr. The values illustrate the regions where velocity shear most actively amplifies the magnetic field along the cold fronts.}
 \label{1600_BB}
\end{figure}

\begin{figure}
 \includegraphics[width=\columnwidth]{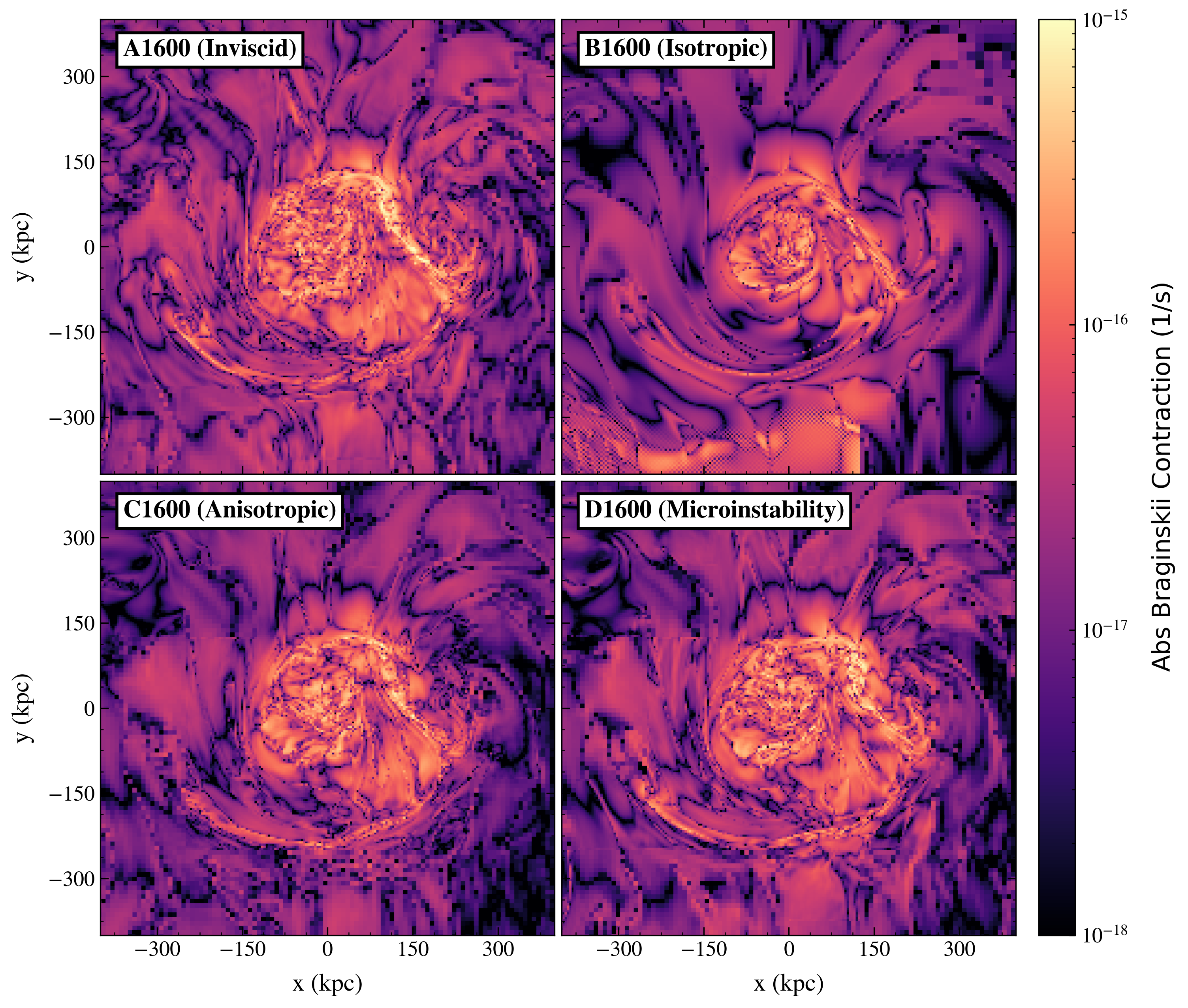}
 \caption{ Slice plots of the Braginskii contraction rate, $|(\vb{b}\vb{b}-\frac{1}{3}\vb{I}):\nabla\bm{v}|$, for the high-$\beta$ ($\beta = 1600$) models at $t = 4.0$ Gyr. This value highlights where the anisotropic viscosity is locally most effective.}
 \label{1600_contraction}
\end{figure}

\begin{figure}
 \includegraphics[width=\columnwidth]{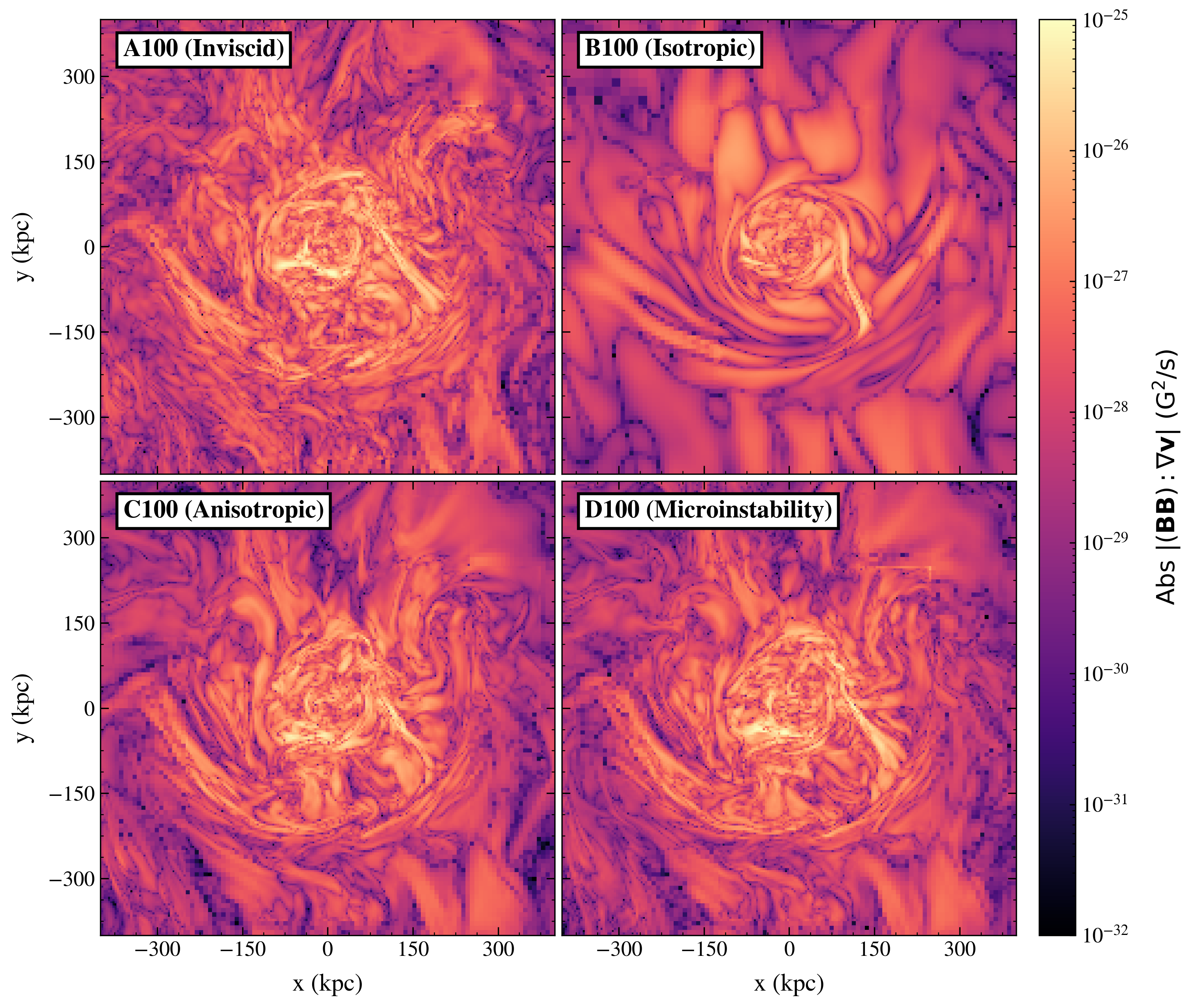}
 \caption{Same as Figure \ref{1600_BB}, but for the strongly magnetized, low-$\beta$ ($\beta = 100$) simulations. Notice the more structured and coherent magnetic stretching compared to the high-$\beta$ cases.}
 \label{100_BB}
\end{figure}

\begin{figure}
 \includegraphics[width=\columnwidth]{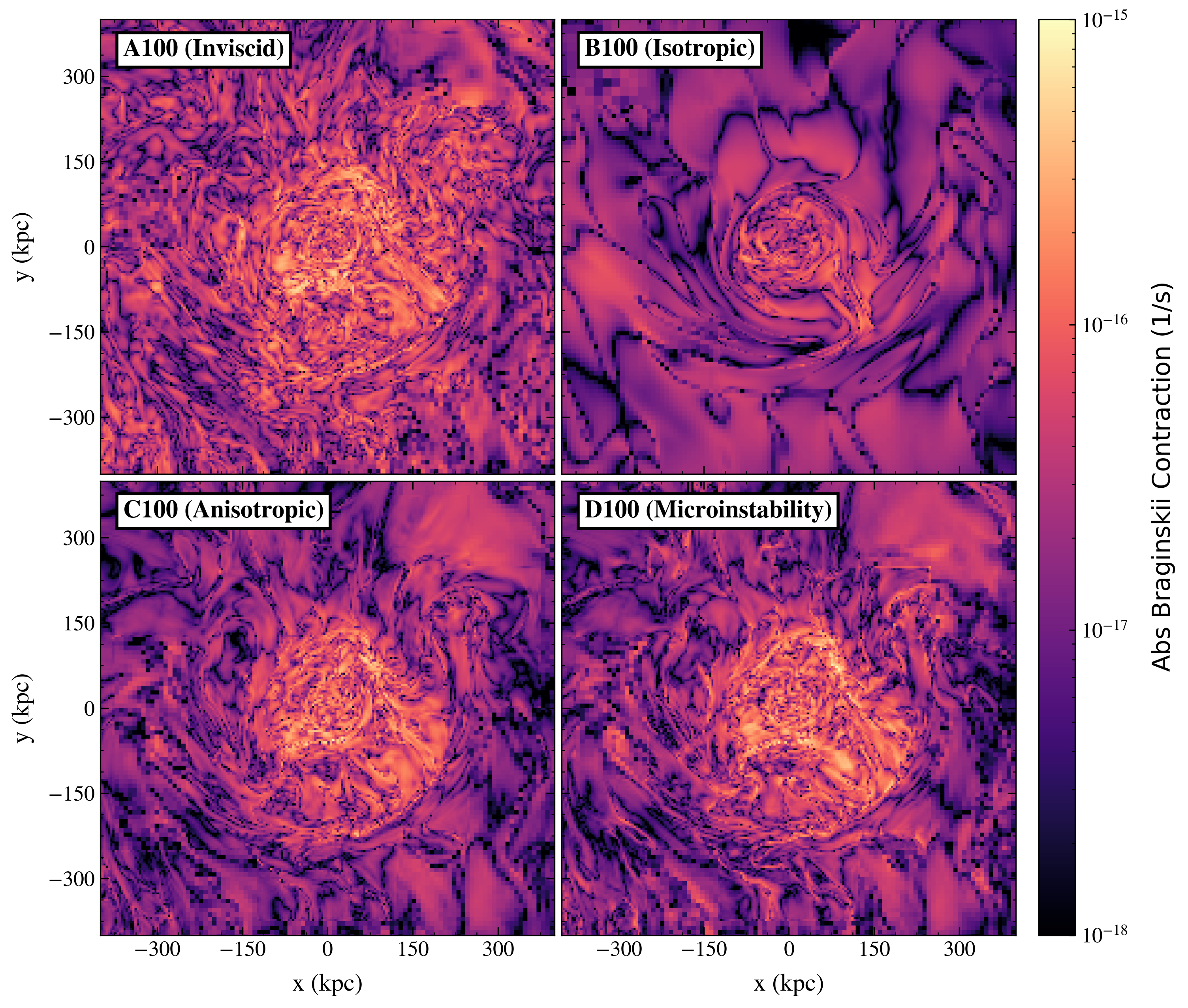}
 \caption{Same as Figure \ref{1600_contraction}, but for the strongly magnetized, low-$\beta$ ($\beta = 100$) simulations. The viscous strain is tightly aligned with the cold front interfaces, further stabilizing the structure against KHI.}
 \label{100_contraction}
\end{figure}

Figure \ref{1600_BB} shows the magnitude of $|(\vb{B}\vb{B}):\nabla\bm{v}|$, which quantifies the stretching of the magnetic field by the local velocity gradient. This diagnostic complements Figure \ref{angle_1600} in the main text by directly identifying the regions where the magnetic field is actively amplified by the sloshing motions. Figure \ref{1600_contraction} presents the Braginskii contraction rate \citep{Braginskii(1965),St-Onge2020},
\[
|(\vb{b}\vb{b}-\tfrac{1}{3}\vb{I}):\nabla\bm{v}|,
\]
where $\vb{b}$ is the unit vector along the magnetic field. This quantity measures the field-aligned rate of strain that enters directly into the Braginskii viscous stress tensor, thereby identifying where anisotropic viscous effects are expected to be strongest. Figures \ref{100_BB} and \ref{100_contraction} show the same diagnostics for the low-$\beta$ simulations. 

In all cases, the cold-front interfaces exhibit enhanced values of $|(\vb{B}\vb{B}):\nabla\bm{v}|$, confirming that magnetic field stretching is concentrated along the sloshing fronts. However, in the low-$\beta$ runs, the stronger magnetic field, corresponding to lower Alfv\'enic Mach numbers ($M_A<2$), resists deformation by the flow through stronger magnetic tension. Consequently, the enhancement of magnetic field stretching across the cold-front interface is less pronounced than in the high-$\beta$ simulations, where magnetic tension is relatively weak ($M_A>5$) and the field can be stretched more efficiently by the sloshing motions.

Comparison with the macroscopic density and temperature distributions reveals a clear morphological contrast in the low-$\beta$ simulations. Although the cold fronts remain smooth and stable owing to the stabilizing effect of magnetic tension, the Braginskii contraction rate (Figure \ref{100_contraction}) exhibits a turbulent, filamentary distribution. This suggests that, while magnetic tension effectively suppresses large-scale KHI roll-ups, the sloshing motions and turbulence continue to generate localized field-aligned velocity-shear structures. As a result, regions of enhanced anisotropic viscous strain remain distributed along the cold-front interface even in the absence of prominent macroscopic instabilities. These additional diagnostics therefore complement the macroscopic analysis by revealing where anisotropic viscous effects are most active during the sloshing evolution.

\bsp	
\label{lastpage}
\end{document}